%% file: main_v2.tex
\documentclass[amssymb,twocolumn,aps,prd,twocolumn,superscriptaddress,floatfix,nopacs,preprintnumbers,nofootinbib]{revtex4-1}
\usepackage[utf8]{inputenc}

\input{packages}
\input{definitions}

\newcommand{\update}[1]{#1}

\begin{document}

\author{Heikki Mäntysaari}
\email{heikki.mantysaari@jyu.fi}
\affiliation{Department of Physics, University of Jyväskylä, P.O. Box 35, 40014 University of Jyväskylä, Finland}
\affiliation{Helsinki Institute of Physics, P.O. Box 64, 00014 University of Helsinki, Finland}

\author{Jani Penttala}
\email{janipenttala@physics.ucla.edu}
\affiliation{Department of Physics and Astronomy, University of California, Los Angeles, CA 90095, USA}
\affiliation{Mani L. Bhaumik Institute for Theoretical Physics, University of California, Los Angeles, CA 90095, USA}

\author{Farid Salazar}
\email{faridsal@uw.edu}
\affiliation{Institute for Nuclear Theory, University of Washington, Seattle WA 98195-1550, USA}

\author{Björn Schenke}
\email{bschenke@bnl.gov}
\affiliation{Physics Department, Brookhaven National Laboratory, Upton, NY 11973, USA}

\title{Finite-size effects on small-$x$ evolution and saturation in proton and nuclear targets}

\begin{abstract}
    Within the Color Glass Condensate effective field theory, we assess the importance of including a finite size for the target on observables sensitive to small-$x$ evolution. 
    To this end, we study the Balitsky--Kovchegov (BK) equation with impact-parameter dependence in the initial condition.
   We demonstrate that neglecting the dependence on the impact parameter can result in overestimated saturation effects for protons, while it has little effect for heavy nuclei at the energies available at current experiments.
    When fixing the nonperturbative parameters to the energy dependence of the exclusive $\jpsi$ photoproduction cross section with proton targets,
   predictions for  lead targets are not sensitive to the applied running-coupling prescription, the scheme chosen to resum large transverse logarithms in the BK equation, 
   or the infrared regulator in the evolution. 
\end{abstract}

\maketitle

\section{Introduction}

At high energies, the partonic structure of protons and nuclei starts to be dominated by gluons.
This gluonic state of matter can be described using the Color Glass Condensate (CGC)~\cite{Iancu:2003xm} effective field theory, which describes the high gluon density regime in terms of classical color fields.
Using the CGC framework, the energy dependence of scattering processes can be described using the JIMWLK~\cite{Jalilian-Marian:1997qno,Jalilian-Marian:1997ubg,Kovner:2000pt,Iancu:2000hn,Mueller:2001uk} or Balitsky--Kovchegov (BK)~\cite{Kovchegov:1999yj,Balitsky:1995ub} equations that naturally include the nonlinear effects of QCD that lead to gluon saturation at high energies.
The saturation effects are expected to eventually slow down the rapid energy dependence of cross sections that would otherwise violate the Froissart--Martin bound \cite{Froissart:1961ux,Martin:1965jj} and thus the unitarity of the scattering \cite{Iancu:2001md,Ferreiro:2002kv}.
While there have been hints of gluon saturation in experimental data \cite{Albacete:2014fwa,Morreale:2021pnn}, no one clear piece of evidence has been identified. It is anticipated that a wide range of experimental observables from different experiments will be needed to establish the presence of gluon saturation below a sufficiently small $x$, the longitudinal momentum fraction of the gluons in the considered hadron or nucleus. 

One  advantage of the CGC framework is the possibility to describe inclusive and diffractive processes using the same degrees of freedom.
In terms of measuring the transverse structure of the target nuclei, diffractive processes are especially powerful as their dependence on the Mandelstam $t$ variable is directly related to the impact-parameter (IP) dependence of the transverse nuclear density by a Fourier transform.
Conversely, this means that to describe the $t$-dependence of diffractive processes one has to take into account the IP dependence of the scattering amplitude.

In phenomenological computations in the CGC framework, IP dependence is often neglected to simplify calculations.
This approximation is valid when the target is much larger than the relevant length scales probed in the scattering process.
However, gluon saturation effects are most prominent when the typical momentum scale probed by the scattering process is comparable to the saturation scale $Q_s$,  which is expected to be moderate, $\mathcal{O}(\SI{1}{GeV})$, at present collider energies.
Especially for protons, the relevant length scales $1/Q_s$ can then be comparable to the proton size, questioning the validity of the approximation in neglecting the impact parameter.
Assessing the importance of including the full IP dependence is one major motivation for this work.

The IP dependence in the CGC framework has been studied previously in several different works, e.g.~in \cite{Berger:2010sh,Berger:2011ew,Berger:2012wx,Schlichting:2014ipa,Mantysaari:2018zdd,Bendova:2019psy,Cepila:2020xol,Mantysaari:2020lhf,Mantysaari:2022sux,Cepila:2023pvh}, incorporating it both in the high-energy evolution describing the scattering off the target and in the initial condition for the evolution equations.
In general, it has been found that including the IP dependence leads to the  Coulomb tails in the IP dependence of the dipole--target scattering amplitude, leading to a rapid growth of the target size that violates the Froissart--Martin bound, rendering it unphysical \cite{Kovner:2001bh}.
This problem arises from splitting into large dipoles and requires 
a description of the
confinement scale effects, which are not included in perturbative calculations.
This problem does not appear when the IP dependence is neglected because then the target is approximated to be infinite and there is no contribution from the target growth.
Understanding the correct treatment of the nonperturbative region in the high-energy evolution is still an open problem.

Another modification to the high-energy evolution that is often considered is the resummation of higher-order corrections~\cite{Beuf:2014uia,Ducloue:2019ezk,Iancu:2015vea,Iancu:2015joa}. 
These appear as large transverse logarithms in the evolution, and it has been demonstrated that their inclusion is crucial for the stability of the BK evolution at next-to-leading logarithmic (NLL) accuracy~\cite{Lappi:2015fma,Lappi:2016fmu}.
There exist multiple different schemes that can be used to resum these corrections, all of which generically result in slower evolution \cite{Beuf:2020dxl}.

In this work, we assess the importance of the impact-parameter dependence on small-$x$ evolution, using the BK equation. This setup allows us to study the effects of the resummation of higher-order corrections and different running-coupling prescriptions in a full IP-dependent framework.  
As a direct application, this allows us to investigate the origin of tensions in the simultaneous description of diffractive vector meson production in $\gamma+p$ and $\gamma+A$ collisions, observed when using the leading-logarithmic evolution. In particular, CGC calculations typically cannot predict as large nuclear suppression as observed at the LHC~\cite{Mantysaari:2022sux,Mantysaari:2023xcu}. Such study is currently not possible within the framework used in Refs.~\cite{Mantysaari:2022sux,Mantysaari:2023xcu} where the JIMWLK evolution equation is used, as currently no numerical implementation that would resum the higher-order corrections in the evolution is available (see however Refs.~\cite{Hatta:2016ujq,Korcyl:2024xph,Korcyl:2024zrf} for related developments), and only a few different running-coupling schemes can be implemented~\cite{Lappi:2012vw,Altinoluk:2023krt}.

This article is organized as follows.
In Sec.~\ref{sec:initial_condition}, we describe the IP-dependent initial condition for the high-energy evolution. The evolution equation along with 
running-coupling corrections and resummation of other higher-order corrections
is described in Sec.~\ref{sec:evolution}. 
The numerical results are presented in Sec.~\ref{sec:numerical_results} where we discuss the effect of including the full IP dependence (Sec.~\ref{sec:IPvsNOIP}), the running-coupling description (Sec.~\ref{sec:results_coupling}), the resummation of higher-order corrections (Sec.~\ref{sec:results_constraint}), and the infrared regulator (Sec.~\ref{sec:regulator}).
Finally, we summarize our conclusions of the results in Sec.~\ref{sec:conclusions}.

\section{Impact-parameter-dependent initial condition}
\label{sec:initial_condition}

To study the effects of including the full IP dependence in the dipole amplitude, we start with a model for the IP-dependent initial condition for the high-energy evolution. We take this initial condition to be a generalization of the McLerran--Venugopalan (MV) model~\cite{McLerran:1993ka,McLerran:1993ni,McLerran:1994vd} by introducing a dependence on the transverse coordinate $\xt$ to the color correlator.
This type of a generalized MV model is very natural for the JIMWLK evolution where it has been used widely as an initial condition, see e.g.~Refs.~\cite{Schenke:2012wb,Schlichting:2014ipa,Mantysaari:2022sux}, and its implications on 2-particle correlators have also been studied in Refs.~\cite{Iancu:2017fzn,Salazar:2019ncp}.
Specifically, we consider
the Gaussian weight 
\begin{equation}
\label{eq:W_MV}
    \mathcal{W}[\rho] = \exp(-\int\dd[2]{\xt}\dd{x^+} \frac{\Tr[ \rho^2(\xt,x^+)]}{\mu^2(\xt,x^+)}) \,,
\end{equation}
where $\rho=\rho^a t^a$ are the color charges, and the color charge density $\mu^2(\xt, x^+)$ now depends on the transverse coordinate.
We also introduce an infrared (IR) regulator $m$ when solving for the gluon field $A^{a-}$ from the classical Yang--Mills equation
\begin{equation}
\label{eq:rho_from_A}
     (-\boldsymbol{\nabla}^2 + m^2) A^{a-} = \rho^a \,.
\end{equation}
This infrared regulator acts effectively like a transverse mass in the gluon propagator that suppresses large Coulomb tails.

With the weight function~\eqref{eq:W_MV} it is now possible to calculate small-$x$ correlators of the light-like Wilson lines
\begin{equation}
    V(\xt) = \mathcal{P} \exp( - i\gs \int_{-\infty}^\infty \dd{x^+} A^-(\xt, x^+) )\,,
\end{equation}
where $\gs=\sqrt{4\pi \as}$ is the strong coupling constant.
The calculation of the correlator of two light-like Wilson lines, corresponding to a scattering of a quark--antiquark dipole, can be done in the same way as for the standard MV model without the impact-parameter dependence~\cite{Gelis:2001da}, and one obtains
\begin{multline}
\label{eq:S_ip}
S(\xt, \yt) 
=\expval{\hat S(\xt, \yt) }
= \frac{1}{N_c}\left\langle  \Tr[V(\xt) V^\dag(\yt)] \right\rangle \\
= \exp(
    - \frac{\as  C_F}{2\pi} \int \dd[2]{\zt}\dd{z^+}   \mu^2(\zt, z^+)  \Gamma_{\zt}(\xt,\yt)  ) \,,
\end{multline}
where $\langle \dots \rangle$ denotes the expectation value with respect to the Gaussian weight \eqref{eq:W_MV} and
\begin{equation}
\label{eq:Gamma}
\Gamma_{\zt}(\xt,\yt) = 
    \bigl[ K_0( m \abs{\xt-\zt} )  -  K_0( m \abs{\yt-\zt} ) \bigr]^2 \,.
\end{equation}
Here the Bessel function $K_0$ appears because it is the Green's function of the differential operator in Eq.~\eqref{eq:rho_from_A}:
\begin{equation}
     (-\boldsymbol{\nabla}^2 + m^2) K_0(m \abs{\xt}) = 2\pi\delta^{(2)}(\xt) \,.
\end{equation}
We shall call this model, specified by Eqs.~\eqref{eq:W_MV} and \eqref{eq:rho_from_A}, the \textit{impact parameter dependent MV} (IP-MV) model.

The dipole amplitude in the  MV model can be recovered from Eq.~\eqref{eq:S_ip} by 
assuming that the target is much larger than the relevant dipole sizes $r =\abs{\rt}$ where $\rt = \xt -\yt$.
This means that the color charge density $\mu^2(\zt, z^+)$ varies  slowly compared to $\Gamma_\zt$ and we can perform the replacement $\mu^2(\zt, z^+) \to \mu^2(\bt, z^+)$ in Eq.~\eqref{eq:S_ip}, where $\bt = \frac{1}{2} (\xt+\yt)$ is the impact parameter.
The $\zt$-integral can then be done analytically, leading to
\begin{equation}
\label{eq:Gamma_integrated}
    \int \dd[2]{\zt} \Gamma_\zt\qty(\xt, \yt)
    = \frac{2\pi}{m^2} \qty[ 1- m r K_1(mr) ] \,,
\end{equation}
and thus
\begin{multline}
\label{eq:S_MV}
S_\text{MV}(\xt, \yt) \\
= \exp(
    - \frac{\alpha_s C_F}{m^2}
    \qty[\int \dd{z^+}   \mu^2(\bt, z^+)  ]
    \qty[ 1- m r K_1(mr) ]
    ) \,.
\end{multline}
We will call this the MV model, although usually, one neglects the dependence on the impact parameter and also expands the term in the exponential around $mr \ll 1$ with
\begin{equation}
    1- m r K_1(mr) \approx \frac{m^2 r^2}{2} \log(\frac{1}{mr}) \,.
\end{equation}

To demonstrate the effects of including the dependence on the impact parameter, we will consider a specific model for the color charge density $\int \dd z^+ \mu^2(\zt,z^+)$.
However, the general features of the dipole amplitude are expected to depend only weakly on the considered model.
In this paper, the model used for the color charge density of the proton is given by
\begin{equation}
\label{eq:IP_model}
     \frac{\as C_F}{2\pi}\int \dd{z^+} \mu^2(\zt, z^+) \equiv  \kappa T(\zt) \,,
\end{equation}
where $\kappa$ is taken to be a constant and $T(\zt)$ is the thickness function of the target normalized such that
\begin{equation}
    \int \dd[2]{\bt} T(\bt) =1 \,.
\end{equation}
Here we ignore the running of the coupling constant and absorb its value to the constant $\kappa$.
For the thickness function $T$ we follow Ref.~\cite{Mantysaari:2022sux} and choose a Gaussian profile for protons:
\begin{equation}
    T(\bt) = \frac{1}{2\pi B_p} e^{-\bt^2/(2B_p)} \,.
\end{equation}
The parameter values describing the proton structure at the initial rapidity $Y_0 = \log 1/0.01$ are chosen as
\begin{align}
    B_p &= \SI{3}{GeV^{-2}}\,,
    &
    m &= \SI{0.4}{GeV}\,,
    &
    \kappa &= \num{0.66}\,,
\end{align}
where the values for the proton size $B_p$ and the infrared cutoff $m$ follow Ref.~\cite{Mantysaari:2022sux}, and
the value of $\kappa$ has been chosen to approximate very closely the initial condition of the small-$x$ evolution  in Ref.~\cite{Mantysaari:2022sux} that has been fitted to exclusive $\jpsi$ production data~\cite{H1:2005dtp,H1:2013okq,ZEUS:2002wfj,ALICE:2014eof,ALICE:2018oyo,LHCb:2014acg, LHCb:2018rcm}. 
We will demonstrate in Sec.~\ref{sec:numerical_results} that the dipole amplitude obtained here results in similar cross sections for $\jpsi$ photoproduction as the one obtained e.g.~in Ref.~\cite{Mantysaari:2022sux} where the MV model is implemented on a transverse lattice.

For nuclear targets, we use the optical Glauber to model the initial condition for the dipole amplitude. 
This only modifies the thickness function in Eq.~\eqref{eq:IP_model} by
\begin{equation}
      T(\bt) \to A T_A(\bt) \,,
\end{equation}
where the nuclear thickness function $T_A$ is given by the  Woods--Saxon distribution integrated over the longitudinal coordinate $z$
\begin{equation}
    T_A(\bt) = \int \dd{z} \rho_A(\bt, z) \,,
\end{equation}
where
\begin{equation}
    \rho_A(\bt, z)
    = \frac{n}{ 1+ \exp[ \frac{\sqrt{\bt^2 + z^2 } - R_A}{d} ]} \,.
\end{equation}
The parameters in $\rho_A$ are chosen as $d = \SI{0.54}\,\si{fm}$ and $R_A = ( 1.12 A^{1/3} - 0.86 A^{-1/3} )\,\si{fm}$, and $n$ is a normalization constant that is determined from 
the condition $\int \dd[2]{\bt} \dd{z} \rho_A(\bt ,z) =1$ as e.g. in Ref.~\cite{Lappi:2013zma}.

\begin{figure}[tb]
    \centering
    \includegraphics[width=\columnwidth]{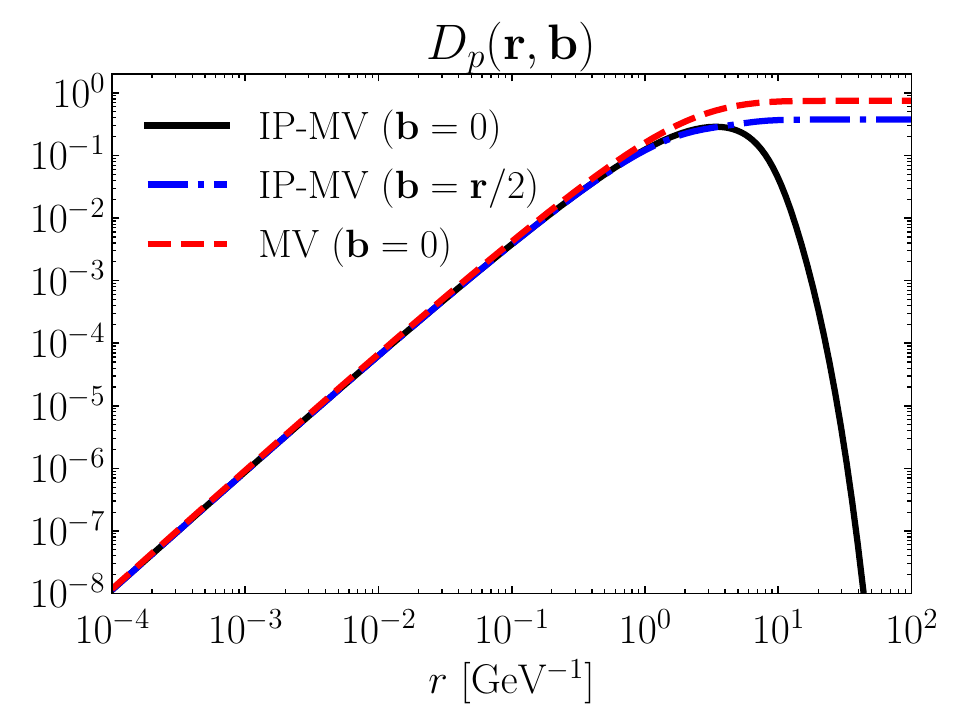}
    \caption{The initial condition for the dipole amplitude as a function of dipole size $r$, using the MV \eqref{eq:S_MV} and IP-MV \eqref{eq:S_ip} models.
    The $\bt=\rt/2$ case refers to the configuration where one of the particles in the dipole is at the origin (the center of the target), and in the $\bt=0$ case the center of the dipole is at the origin.
    }
    \label{fig:dipole_amplitude_comparison_fixed_b}
\end{figure}

While the MV model~\eqref{eq:S_MV} is valid for dipole sizes much smaller than the target size, the behavior for large dipoles is wildly different compared to the impact-parameter-dependent case (Eq.~\eqref{eq:S_ip}).
This is illustrated in Fig.~\ref{fig:dipole_amplitude_comparison_fixed_b} where we show the dipole amplitude
\begin{equation}
    D(\rt, \bt) = 1 - S(\xt, \yt) \,,
\end{equation}
computed using the IP-MV~\eqref{eq:S_ip} and MV~\eqref{eq:S_MV} models as a function of the dipole size $r = \abs{\rt}=\abs{\xt-\yt}$.
We show two different configurations for the IP-MV model:
\begin{enumerate}
    \item\label{it:ip_at_origin} The case where the impact parameter $\bt$ 
    is set to the center of the target ($\bt=0$).
    In this configuration, both quarks in the dipole go further away from the target as the dipole size $r$ increases.
    \item\label{it:quark_at_origin} The case where the location of one of the quarks is fixed to the center of the target while the other is located at a distance $r$ away from the target ($\bt=\rt/2$).
\end{enumerate}
For small dipole sizes ($r\ll$ target size) these two configurations agree with each other as well as the original MV model, whereas for large $r$ the behavior is very different.
The different large-$r$ behavior can be explained by the observation that in our model the asymptotic behavior for the Wilson lines is $V(\xt) \to 1$ when $\abs{\xt} \to \infty$ (recall that far away from the target $A^-=0$). 
Doing this substitution for the dipole amplitude with Configuration~\ref{it:ip_at_origin} makes it vanish identically, whereas Configuration~\ref{it:quark_at_origin} reduces to 
{
\begin{equation}
   D(\rt,\bt =\rt/2)
    \overset{\abs{\rt} \to \infty}{\approx}
    1 - \frac{1}{\Nc} \expval{\Tr V(0)}
\end{equation}
which is non-zero~\cite{Gelis:2001da}.}
For a fixed impact parameter (e.g.~Configuration 1), the IP-MV and MV models also differ most at large $r$,
as for the IP-MV model the asymptotic behavior at  $r\rightarrow \infty$ is $D(\rt, \bt) \to 0$ compared to $D(\rt, \bt) \to 1$ for the MV model. 
In other words, these two models include an opposite effective description of confinement scale physics.

\section{High-energy evolution of the dipole amplitude}
\label{sec:evolution}

\subsection{Balitsky--Kovchegov evolution}

The dipole amplitude depends on the energy of the dipole--target scattering, and this energy dependence can be computed using the perturbative JIMWLK evolution equation. It can be written as a Fokker--Planck equation describing the energy dependence of the weight $\mathcal{W}$ in Eq.~\eqref{eq:W_MV}, or alternatively as a B-JIMWLK hierarchy of equations coupling the evolution of different $n$-point correlators of Wilson lines~\cite{Balitsky:1995ub}. 
In practice, the JIMWLK evolution is implemented using a Langevin formulation to evolve individual Wilson lines instead of the weight function~\cite{Weigert:2000gi,Blaizot:2002np,Cali:2021tsh}, see also Appendix~\ref{app:JIMWLK}.
If one assumes that the correlators of several dipole operators factorize, i.e.~
$ \expval*{ \hat S  \hat S } \approx  \langle \hat  S \rangle \langle \hat  S \rangle$ which is appropriate in the limit of a large number of colors or in the mean-field limit, 
one can simplify the JIMWLK equation to a closed differential equation written only in terms of the dipole amplitude.
This results in the Balitsky--Kovchegov (BK) equation
\begin{equation}
    \label{eq:BK}
    \partial_Y S(\xt_0, \xt_1,Y)
    = \int \dd[2]{\xt_2} K_\text{BK} (\xt_0,\xt_1,\xt_2) W(\xt_0,\xt_1,\xt_2) \,,
\end{equation}
where $K_\text{BK}$ is the BK kernel and $W$ contains the parts related to the Wilson lines, given by
\begin{multline}
    W(\xt_0,\xt_1,\xt_2)
    = S(\xt_0,\xt_2,Y) S(\xt_2,\xt_1,Y) \\
    - S(\xt_0,\xt_1,Y) \,.
\end{multline}
The BK kernel can be naturally written as 
\begin{multline}
    \label{eq:BK_kernel}
     K_\text{BK} (\xt_0,\xt_1,\xt_2) 
     = \frac{\alpha_s N_c}{2\pi^2} \\
     \times \qty[ \Kt^i(\xt_{20}) - \Kt^i(\xt_{21}) ] \qty[ \Kt^i(\xt_{20}) - \Kt^i(\xt_{21}) ]\,,
\end{multline}
{where $\xt_{ij} = \xt_i - \xt_j$. The vector $\Kt^i(\xt)$ describes
the emission of a soft gluon\footnote{Here ``soft'' refers to a particle carrying a small longitudinal momentum fraction of the emitting particle's momentum.}
with a transverse separation $\xt$ from the emitting (anti)quark}:
\begin{equation}
\label{eq:BK_kernel_regulator}
\Kt^i(\xt) = m'|\xt| K_1(m'|\xt|) \frac{\xt^i}{\xt^2}\,.
\end{equation}
Here we have followed  Refs.~\cite{Schlichting:2014ipa,Mantysaari:2018zdd} and introduced an infrared regulator \update{$m'>0$} that suppresses the long-distance Coulomb tails that would otherwise result in rapidity evolution violating the Froissart--Martin bound~\cite{Kovner:2001bh,Schlichting:2014ipa,Golec-Biernat:2003naj}.
In the limit $m'\to 0$ this reduces to
\begin{equation}
    \Kt^i(\xt) = \frac{\xt^i}{\xt^2} \,,
\end{equation}
which corresponds to the standard BK equation without the IR regulator.

We choose the value $m' = m =\SI{0.4}{GeV}$  for the infrared regulators
as our standard setup following Ref.~\cite{Mantysaari:2022sux} where these parameters are constrained by the exclusive $\jpsi$ production data.
Our setup for the evolution 
closely matches that of Ref.~\cite{Mantysaari:2022sux} with  the color charge density specified in Eq.~\eqref{eq:IP_model}.
Note that in general $m$ and $m'$, corresponding to the IR regulators in the initial condition and the evolution respectively,
are allowed to differ
even though both can be understood as an ``effective'' transverse mass of the gluon. 
Sensitivity to the IR regulator $m'$ will quantified in Sec.~\ref{sec:regulator} in more detail. 
The effect of the regulator $m$ in the initial condition has been studied e.g. in Ref.~\cite{Mantysaari:2016jaz} in the context of exclusive vector meson production.

\subsection{Running coupling in the BK evolution}
\label{sec:running_coupling}

The running of the coupling constant is given by
\begin{equation}
\label{eq:strong_coupling}
    \alpha_s(r^2)
    = \frac{4 \pi}{\beta_0 
    \log \qty[   \qty(\frac{\mu_0^2}{\lambdaQCD^2})^{1 / \zeta}+
     \qty(\frac{4}{r^2\lambdaQCD^2})^{1 / \zeta}]^\zeta
    } \,,
\end{equation}
where $\beta_0 = (11N_c - 2 N_f)/3 $.
We choose the number of flavors $N_f = 3$, and the infrared region is regulated with the parameters
$\mu_0 = \SI{0.28}{GeV}$, and $\zeta= 0.2$.
The value of $\lambdaQCD = \SI{0.025}{GeV}$ is chosen to reproduce the correct energy dependence of the exclusive $\jpsi$ production data,  in agreement with studies using the JIMWLK evolution~\cite{Mantysaari:2022sux}.

For the running coupling in the evolution, we need to choose the scale $r^2$, which is not unique.
In this work we will consider the following prescriptions:
\begin{enumerate}
    \item \textit{Daughter dipole}:
    The scale for $\alpha_s$ is set by the two daughter dipoles, which is achieved by rewriting all terms in the BK kernel as 
    \begin{equation}
     \alpha_s \Kt^i(\xt) \Kt^i(\yt) \to 
    \sqrt{\alpha_s(\xt^2)} \Kt^i(\xt)  \sqrt{\alpha_s(\yt^2)}\Kt^i(\yt)\,.
    \end{equation}
    \item \textit{Parent dipole}:
    The scale is always set by the size of the parent dipole, i.e., $r^2 = \xt_{01}^2$.
    \item \textit{Smallest dipole}:
    The scale is always the smallest of the dipoles, i.e.,
    $r^2 = \min( \xt_{01}^2, \xt_{20}^2, \xt_{21}^2 )\,.$
    \item \textit{Balitsky prescription}:
    The BK kernel~\eqref{eq:BK_kernel} is modified to~\cite{Balitsky:2006wa}
    \begin{multline}
         K_\text{BK} (\xt_0,\xt_1,\xt_2) 
     = \frac{ \alpha_s(\xt_{01}^2) N_c}{2\pi^2} \\
     \times \Biggl\{ \qty[\Kt^i(\xt_{20}) - \Kt^i(\xt_{21}) ] \qty[ \Kt^i(\xt_{20}) - \Kt^i(\xt_{21})] \\
     +\Kt^2(\xt_{20}) \qty( \frac{\alpha_s(\xt^2_{20})}{\alpha_s(\xt^2_{21})} -1)
     +\Kt^2(\xt_{21}) \qty( \frac{\alpha_s(\xt^2_{21})}{\alpha_s(\xt^2_{20})} -1)
     \Biggr\} \,.
    \end{multline}
    Note that our inclusion of the IR regulator modifies the original Balitsky prescription slightly, but in the limit $m'\to 0$ it reduces to the original version.
\end{enumerate}
We choose the daughter dipole prescription as our default setup.
The main reason for this is that this allows us to make more direct comparisons to the JIMWLK evolution used e.g.~in Ref.~\cite{Mantysaari:2022sux}, where the daughter dipole is a natural choice (see also Ref.~\cite{Lappi:2012vw,Altinoluk:2023krt} for alternative prescriptions), and in particular the parent dipole size is not accessible in JIMWLK, see Eq.~\eqref{eq:jimwlk}.
It has also been argued that the daughter dipole prescription is more physical in the sense that the other prescriptions mentioned above do not lead to a positive definite Hamiltonian when uplifted to the JIMWLK evolution~\cite{Kovner:2023vsy, Altinoluk:2023krt}.

\subsection{Resummation of higher-order corrections}
\label{sec:kinematic_constraint}

Besides the ability to study different running-coupling prescriptions, one big advantage of considering the BK evolution over the full JIMWLK evolution is the possibility of testing different prescriptions for resumming large transverse logarithms that appear at higher orders in perturbation theory.
Their origin lies in the correct time-ordering of subsequent emissions of soft gluons which can be imposed by demanding that the momenta of the gluons is ordered in the target rapidity.
While such corrections also exist for the JIMWLK evolution~\cite{Hatta:2016ujq}, they are much more challenging to implement in the Langevin form needed for the numerical evaluation~\cite{Korcyl:2024xph,Korcyl:2024zrf}.

For the resummation of the higher-order corrections in the BK evolution, we consider three different versions and follow the terminology introduced in Ref.~\cite{Beuf:2020dxl}.
First, we consider the \textit{kinematically constrained} BK (KCBK)~\cite{Beuf:2014uia} which resums anticollinear logarithms from the region $\xt_{20}^2, \xt_{21}^2  \gg \xt_{01}^2$.
Here the Wilson line part is modified to
\begin{multline}
    W(\xt_0,\xt_1,\xt_2)
    = 
    \theta(Y - \Yif - \Delta_{012})\\
    \times \bigl[S(\xt_0,\xt_2,Y -\Delta_{012}) S(\xt_2,\xt_1,Y-\Delta_{012}) \\
    - S(\xt_0,\xt_1,Y)\bigr] \,,
\end{multline}
where
\begin{equation}
    \Delta_{012}
    = \max \qty{ 0, \log \frac{\min \qty{\xt_{02}^2,\xt_{12}^2}}{\xt_{01}^2} } \,.
\end{equation}
The initial rapidity is chosen as $\Yif = \log 1/0.01$.

{
Second, the \textit{target} BK (TBK)~\cite{Ducloue:2019ezk} is given in terms of the target rapidity
$\eta$.
The TBK equation is written in terms of Wilson line correlators that depend on the target rapidity, and we denote these by
\begin{equation}
 \bar S\qty(\xt ,\yt, \eta =
 Y + \log \qty[\min \qty{1, \abs{\xt-\yt}^2 Q_0^2}]
 )
\end{equation} 
where $Q_0^2 \equiv \SI{1}{GeV}^2$ is the nucleon mass. 
}
This kind of evolution resums collinear logarithms from the region $\xt_{20}^2, \xt_{21}^2  \ll \xt_{01}^2$.
The Wilson line part is then given by
\begin{multline}
    W(\xt_0,\xt_1,\xt_2)
    = 
    \theta(\eta - \etaif - \delta)\\
    \times \bigl[\bar S(\xt_0,\xt_2,\eta -\delta_{02}) \bar S(\xt_2,\xt_1,\eta-\delta_{12}) \\
    - \bar S(\xt_0,\xt_1,\eta)\bigr] \,,
\end{multline}
where
\begin{equation}
    \delta_{kl} = \max \qty{ 0, \log \frac{\xt_{01}^2}{\xt_{kl}^2} } \,,
\end{equation}
and $\delta = \max \qty{ \delta_{02}, \delta_{12} }$.
As before, we choose the initial rapidity as $\etaif = \log 1/0.01$.

Third, we consider a resummation of single and double transverse logarithms that is local in rapidity~\cite{Iancu:2015vea,Iancu:2015joa}. This prescription resums both collinear and anticollinear logarithms by modifying the BK kernel by
\begin{equation}
    K_\text{BK} \to K_\text{BK} K_\text{STL}K_\text{DLA}  \,,
\end{equation}
where $ K_\text{STL}$ and $K_\text{DLA} $ are related to single and double transverse logarithms, respectively.
The single transverse logarithm term reads
\begin{equation}
    K_\text{STL} =\exp( -
    \frac{\alpha_s(\xt_{01}^2) N_c A_1}{\pi}
    \abs{ \log \frac{\xt_{01}^2}{\min \qty{ \xt_{02}^2, \xt_{21}^2 }} }
    ) \,,
\end{equation}
with $A_1 = \frac{11}{12}$
and resums contributions
related to the DGLAP evolution.
The resummation of double transverse logarithms is done with
\begin{equation}
\label{eq:KDLA}
    K_\text{DLA}
     = \frac{J_1 (2 \sqrt{\overline \alpha_s x^2})}{\sqrt{\overline \alpha_s x^2}} \,,
\end{equation}
where $x^2 = \log(\xt_{20}^2/\xt_{01}^2)\log(\xt_{21}^2/\xt_{01}^2)$ and 
$\overline \alpha_s = \alpha_s(\xt_{01}^2) N_c / \pi$.
If $x^2 < 0$, this is analytically continued by substituting $J_1 \to I_1$ and taking the absolute value of $x^2$.
We shall call this version the \textit{resummed} BK (ResumBK). 
As shown in Refs.~\cite{Lappi:2016fmu,Hanninen:2021byo}, this equation can also be used to approximate the NLL BK equation~\cite{Balitsky:2008zza}.
It has also been shown in Ref.~\cite{Bendova:2019psy} that the resummed evolution suppresses the generation of long-distance Coulomb tails at large impact parameters, which can be expected to limit the sensitivity to the regulator $m'$ in the BK kernel~\eqref{eq:BK_kernel_regulator}.

\begin{figure*}[t!]
	\centering
    \begin{subfigure}[T]{0.49\textwidth}
        \centering
        \includegraphics[width=\textwidth]{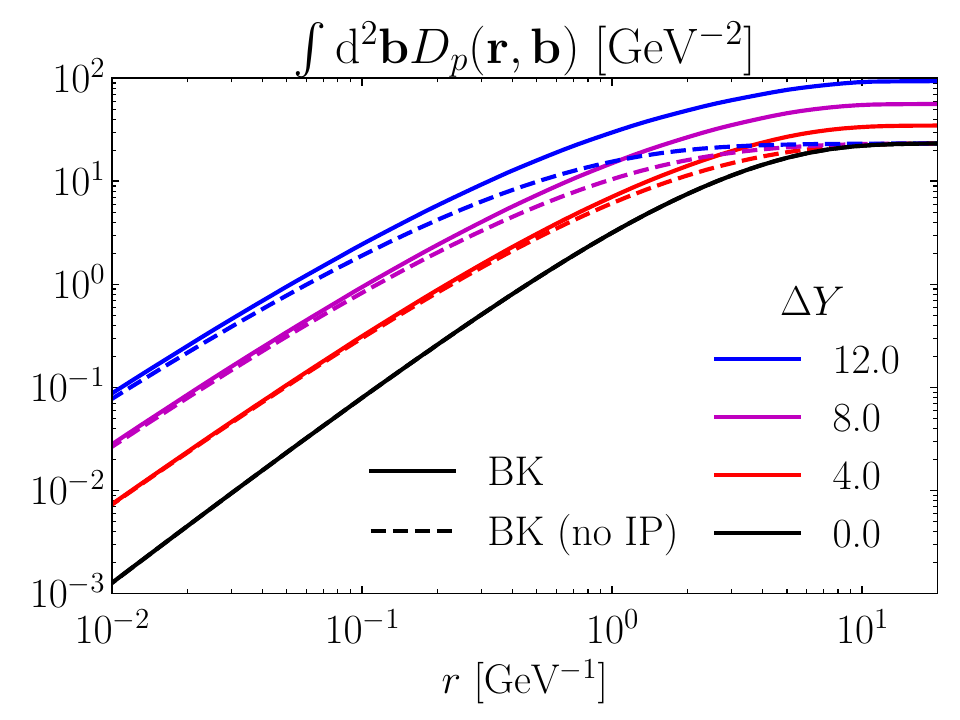}
        \caption{ Proton. }
        \label{fig:evolved_D_ip_vs_noip_p}
    \end{subfigure}
    \begin{subfigure}[T]{0.49\textwidth}
        \centering
        \includegraphics[width=\textwidth]{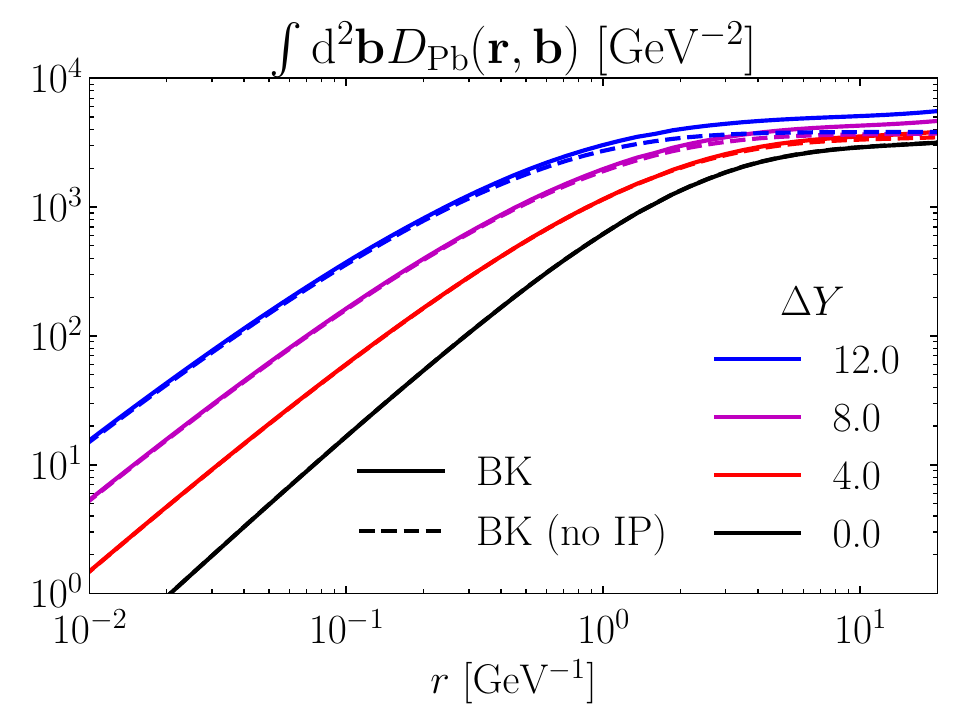}
        \caption{ Lead. }
        \label{fig:evolved_D_ip_vs_noip_Pb}
    \end{subfigure}
    \caption{The IP-integrated dipole amplitude $\int \dd[2]{\bt} D(\rt,\bt)$ evolved using the BK equation with and without dependence on the impact parameter in the initial condition.
    Solid line: with IP dependence in the initial condition. 
    Dashed line: without IP in the initial condition.
    }
    \label{fig:evolved_D_ip_vs_noip}
\end{figure*}

\section{Numerical results}
\label{sec:numerical_results}

\subsection{Impact-parameter dependence in the evolution}

\label{sec:IPvsNOIP}

First, we study the effect of including the full impact-parameter dependence of the dipole amplitude in the evolution. 
In most phenomenological applications, the impact-parameter dependence of the dipole amplitude is neglected and it is assumed to depend only on the dipole size,
which can be effectively understood as considering an IP-averaged dipole amplitude in the evolution.
Thus, to study the effects of neglecting the impact parameter in the evolution, we define the IP-averaged dipole amplitude $\widetilde D$ as
\begin{equation}
\label{eq:D_integrated}
    \int \dd[2]{\bt} D(\rt,\bt) =\sigma_{q}  \widetilde D(r) \,,
\end{equation}
where $\sigma_q$ is a normalization constant that is defined by demanding that $\widetilde D(r) \to 1$ for $r \to \infty$. 
We note that by the optical theorem, the IP-integrated dipole amplitude~\eqref{eq:D_integrated} corresponds to one half of the total dipole--target scattering cross section for a fixed-sized dipole.
We can also give a physical meaning to the constant $\sigma_q$ appearing in Eq.~\eqref{eq:D_integrated} as the total quark--target scattering cross section. 
This follows from the fact that for large dipoles only one of the particles effectively interacts with the target, and by integrating over the impact parameter we 
get a contribution from both the quark and the antiquark scattering separately off the target.
Thus, in the limit $r \to \infty$ Eq.~\eqref{eq:D_integrated} reduces to the sum of the quark--target and the antiquark--target scattering amplitudes, and
by the optical theorem this equals the total quark--target scattering  cross section.

We can then estimate the effects of the impact parameter dependence in the evolution by evolving the average dipole amplitude $\widetilde D$ instead of the full IP-dependent dipole amplitude $D$.
Because of the nonlinear terms in the BK evolution,
the impact parameter integration and the evolution do not commute and there is a difference
whether one takes the IP average before or after the evolution.
This can be seen in Fig.~\ref{fig:evolved_D_ip_vs_noip} where we compare the following two cases:
\begin{enumerate}
    \item First evolve the full IP-dependent dipole amplitude with the BK evolution and then integrate over the impact parameter (denoted by ``BK'' in Fig.~\ref{fig:evolved_D_ip_vs_noip}).
    \item First compute the IP-averaged dipole amplitude and then evolve that with the BK evolution (denoted by ``BK (no IP)'' in Fig.~\ref{fig:evolved_D_ip_vs_noip}).
\end{enumerate}
{
We note that in the dilute limit, where one can neglect the nonlinear term in the BK evolution
and it reduces to the linear Balitsky--Fadin--Kuraev--Lipatov (BFKL) evolution~\cite{Lipatov:1976zz,Kuraev:1977fs,Balitsky:1978ic}, the order of the IP integration and the high-energy evolution does not matter and the two cases lead to the exactly same results.
} 

The two approaches are very close to each other when the dipole size is small compared to the target size. This is because the approximation in neglecting the impact parameter in the evolution essentially assumes that the target varies very slowly compared to the dipole size, making the evolution at different impact parameters roughly independent of each other.
As the dipole size grows, this approximation breaks down and one has to take the impact parameter into account in the evolution.
Most crucially, this affects the behavior of the nonlinear term which behaves very differently for the IP-averaged and unintegrated dipole amplitudes.
As we have demonstrated in Fig.~\ref{fig:dipole_amplitude_comparison_fixed_b}, for a fixed {impact parameter} the unintegrated dipole amplitude approaches zero as the {dipole size} increases, suppressing the nonlinear term in the evolution for large dipoles.
This is the opposite effect compared to the IP-integrated dipole amplitude (i.e. $\widetilde D$ defined in Eq.~\eqref{eq:D_integrated}) that approaches one, enhancing the contribution of the nonlinear term and  ensuring saturation effects in the evolution for large enough dipoles.

Because of this difference in the behavior of the nonlinear term, neglecting the IP dependence in the evolution can lead to overestimating saturation effects.
This is more important for proton targets where the infinite-target approximation breaks down for smaller dipoles, and numerically large differences between the two approaches appear at dipoles $r \lesssim \SI{1}{GeV^{-1}}$ already after a few units of evolution in rapidity. 
This means that using the IP-independent approximation for protons can make a difference in phenomenological applications.
For the Pb nucleus, the situation is very different, as due to its larger size the IP-independent approximation is good over a wider range of dipole sizes. 

\begin{figure*}[t!]
    \begin{subfigure}[T]{0.49\textwidth}
        \centering
        \includegraphics[width=\textwidth]{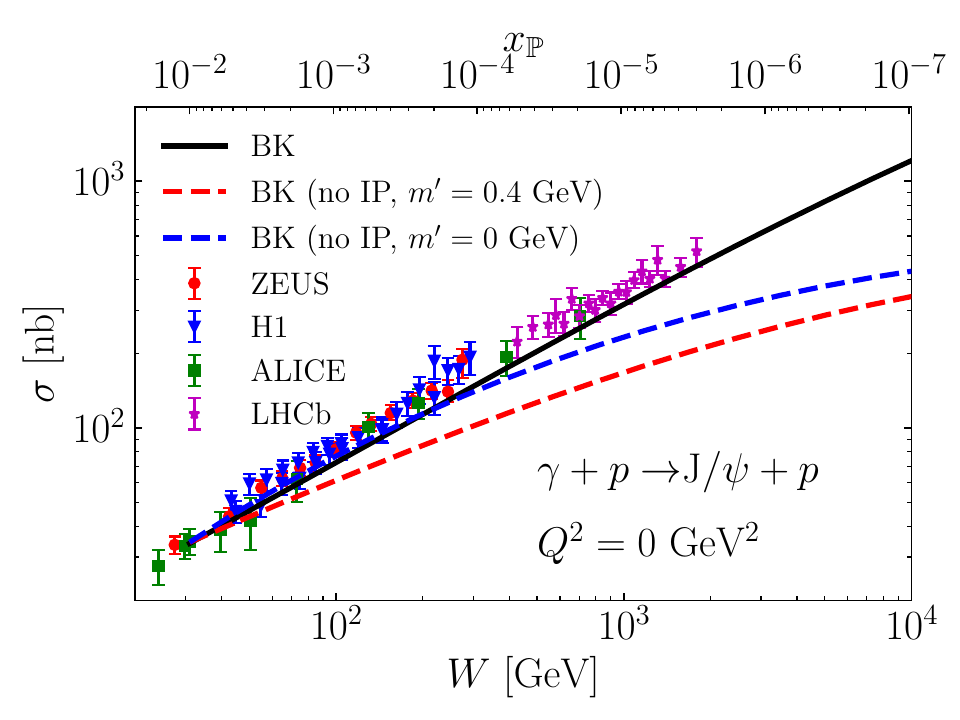}
        \caption{ Proton.}
        \label{fig:exclusive_jpsi_p_ip_vs_noip_p}
    \end{subfigure}
    \begin{subfigure}[T]{0.49\textwidth}
        \centering
        \includegraphics[width=\textwidth]{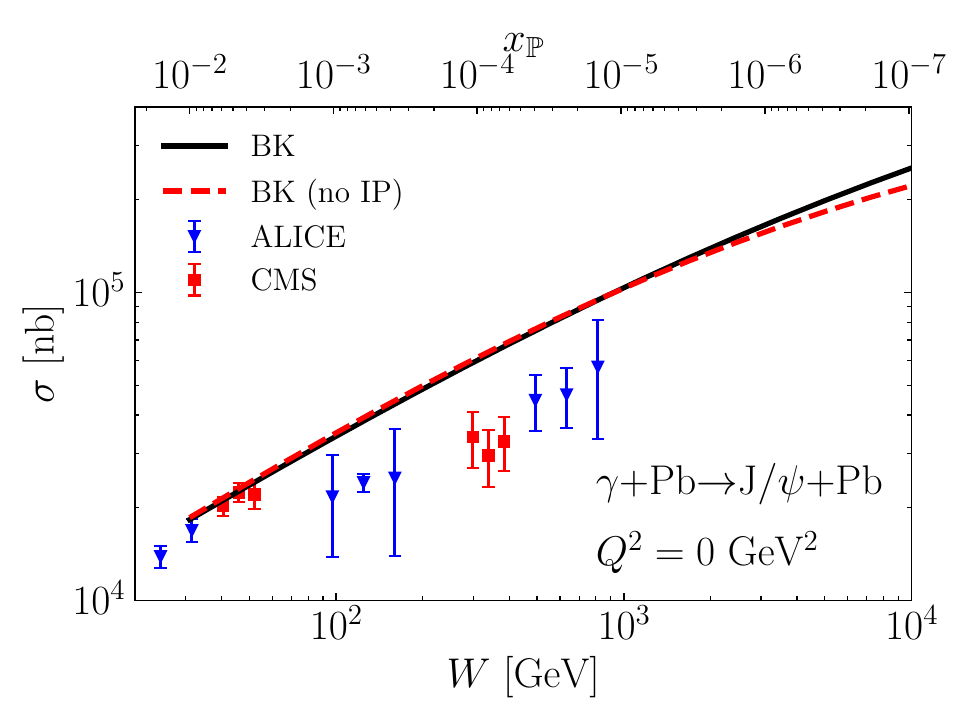}
        \caption{Lead. }
        \label{fig:exclusive_jpsi_p_ip_vs_noip_Pb}
    \end{subfigure}
    \caption{Exclusive $\jpsi$ production as a function of the center-of-mass energy $W$ with and without IP dependence in the initial condition for the BK evolution of the dipole amplitude, compared to the experimental data for proton  (ZEUS~\cite{ZEUS:2002wfj}, 
    H1~\cite{H1:2005dtp,H1:2013okq}, 
    ALICE~\cite{ALICE:2014eof,ALICE:2018oyo}, 
    LHCb~\cite{LHCb:2014acg,LHCb:2018rcm})
    and lead
    (ALICE~\cite{ALICE:2023jgu},
    CMS~\cite{CMS:2023snh})
    targets.
    For protons, we also show the IP-independent case with $m' = \SI{0}{GeV}$ that happens to agree more closely with the energy of dependence of the data.
    }
    \label{fig:exclusive_jpsi_p_ip_vs_noip}
\end{figure*}

{
The different behavior of the IP-averaged and unintegrated dipole amplitudes at large dipoles also explains the difference at large $r$ ($\gtrsim$ target size) for the IP-integrated dipole amplitude after evolution, as shown in Fig.~\ref{fig:evolved_D_ip_vs_noip}.
For the IP-averaged dipole amplitude $\widetilde D(r)$, we have $\widetilde D(r) \approx 1$ at large dipoles so that $\widetilde D$ is saturated and there is no evolution (recall that $\widetilde D(r) = 1$ is a stable fixed point of the BK equation).
For the IP-dependent case, on the other hand, large dipoles correspond to the case where the quark and the antiquark scatter independently off the target.
This scattering, corresponding to the quark--target cross section $\sigma_q$ after the integration over the impact parameter, becomes enhanced with increasing energy due to the increase in the saturation scale and the growth of the target size.
Thus, the value of the integrated dipole amplitude increases with energy even for large dipoles when the IP dependence is included in the evolution.
}

\begin{figure}[t!]
    \centering
    \begin{subfigure}[T]{0.9\columnwidth}
        \centering
        \includegraphics[width=\textwidth]{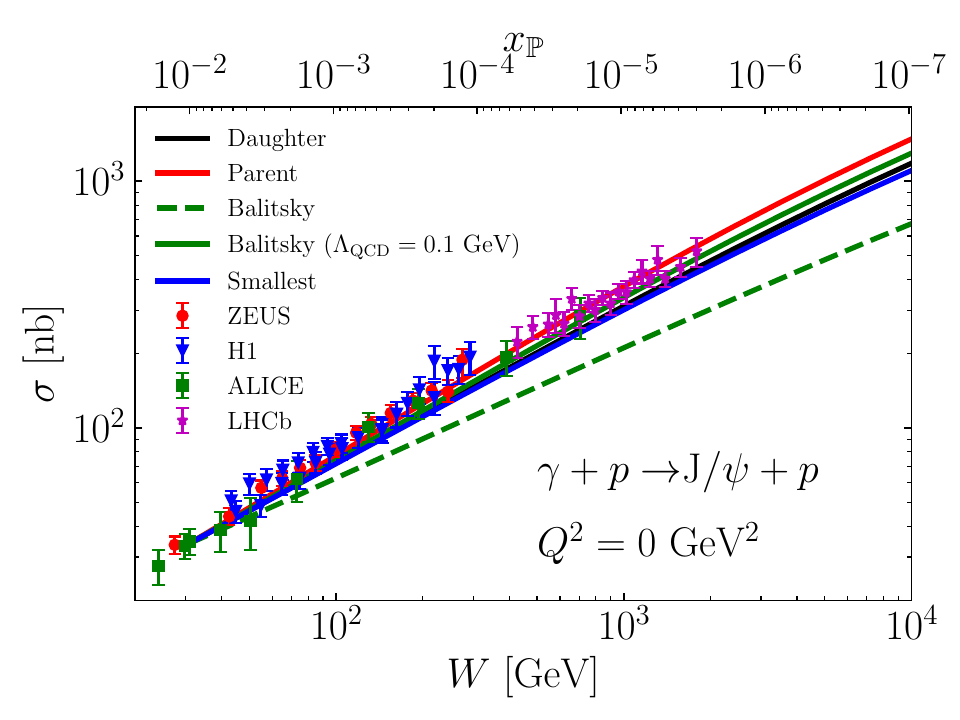}
        \caption{ Proton. }
        \label{fig:exclusive_jpsi_alphas_p}
    \end{subfigure}
    \begin{subfigure}[T]{0.9\columnwidth}
        \centering
        \includegraphics[width=\textwidth]{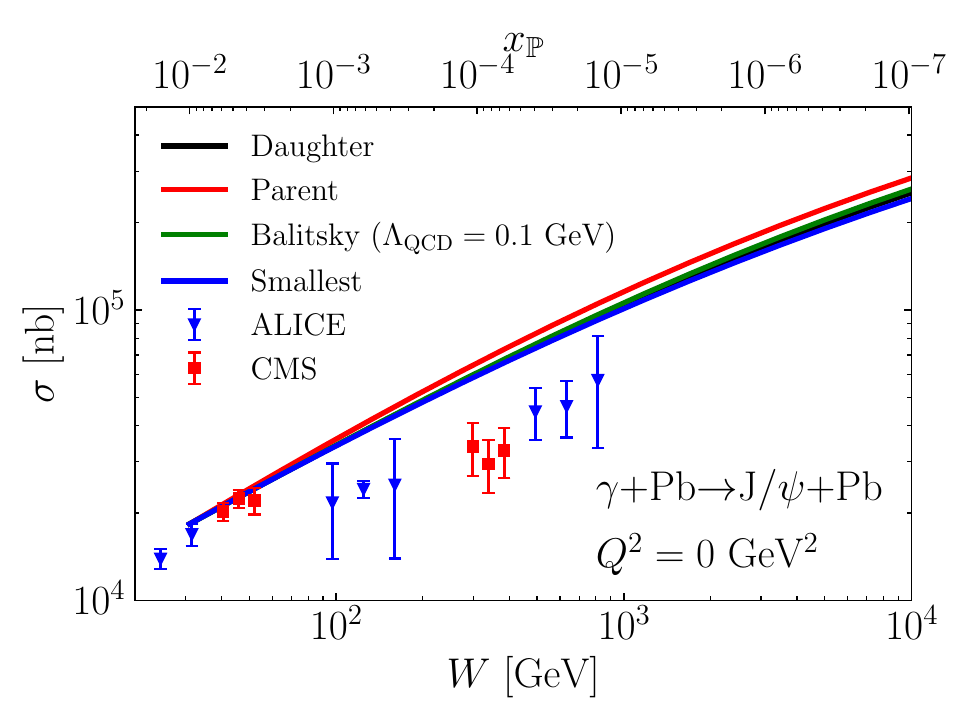}
        \caption{ Lead. }
        \label{fig:exclusive_jpsi_alphas_Pb}
    \end{subfigure}
    \begin{subfigure}[T]{0.9\columnwidth}
        \centering
        \includegraphics[width=\textwidth]{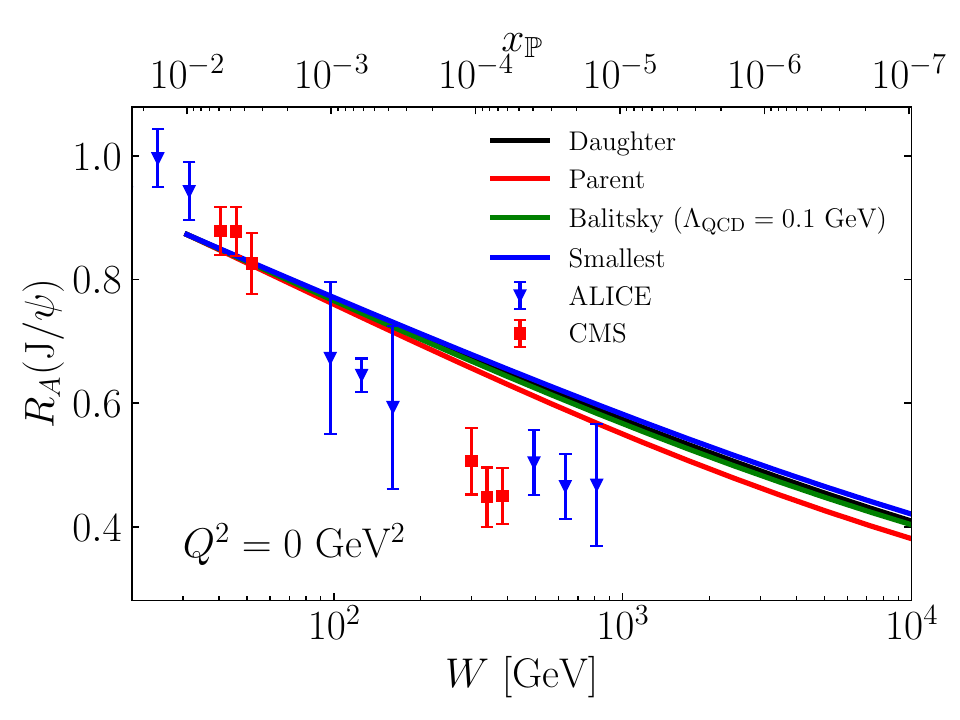}
        \caption{ Nuclear suppression factor.}
        \label{fig:exclusive_jpsi_alphas_R}
    \end{subfigure}
    \caption{
    Exclusive $\jpsi$ production cross section  with
    different prescriptions for the running of the coupling in the BK equation.
    The $\gamma+p$ data are from Refs.~\cite{H1:2005dtp,H1:2013okq,ZEUS:2002wfj,ALICE:2014eof,ALICE:2018oyo,LHCb:2014acg, LHCb:2018rcm}.
    The $\gamma+\mathrm{Pb}$ cross section and the nuclear suppression factor are compared to the
    experimental data from
    ALICE~\cite{ALICE:2023jgu}
    and
    CMS~\cite{CMS:2023snh}.
    }
    \label{fig:exclusive_jpsi_alphas}
\end{figure}

To quantify the phenomenological implications of the IP-independent approximation, we compute exclusive $\jpsi$ photoproduction using the dipole amplitudes obtained with and without the IP dependence in the initial state of the evolution.
We choose to study vector meson photoproduction due to its strong sensitivity to saturation effects and on target geometry; implications on the total DIS cross section are quantified in Appendix~\ref{app:DIS}.
The differential cross section can be written as~\cite{Kowalski:2006hc,Hatta:2017cte,Mantysaari:2020lhf,Marquet:2010cf}
\begin{equation}
    \frac{\dd{\sigma}^{\gamma + p/A \to V + p/A}}{\dd{t}} = \frac{1}{4\pi}  
    \frac{1}{2}\sum_{\lambda = \pm 1} \abs{ \mathcal{A}_\lambda}^2
\end{equation}
where the invariant amplitude is
\begin{multline}
\label{eq:amplitude}
    -i \mathcal{A}_\lambda = \int \dd[2]{\bt} \dd[2]{\rt} \int_0^1 \frac{\dd{z}}{4\pi }
    e^{-i \Deltat \vdot \qty( \bt + (z-\frac{1}{2} ) \rt )}
    \\ 
    \times
     D\qty(\rt, \bt, \Ypom) 
    \psi_\lambda^{\gamma \to q \bar q}(\rt, z) \qty[\psi_\lambda^{V\to q \bar q}(\rt, z)]^*  \,.
\end{multline}
Here the Mandelstam $t$ variable can be written in terms of the transverse-momentum transfer $\Deltat$ as $t \approx - \Deltat^2$, 
{and $\Ypom = \log 1/\xpom$ with}
\begin{equation}
    \xpom = \frac{Q^2 + M_V^2 - t}{ W^2 + Q^2 + m_N^2}
\end{equation}
where $W$ is the center-of-mass energy of the photon--nucleon system, $Q^2$ is the photon virtuality (zero in this case), $M_V$ is the $\jpsi$ mass and $m_N$ is the mass of a nucleon.
We neglect the $t$-dependence in $\xpom$ to simplify calculations, as the differential cross section for coherent $\jpsi$ production is highly peaked at values $t\approx 0$.
{The initial condition in the BK evolution is chosen as the IP-MV model at the rapidity $Y_0 = \log 1/0.01$.}
The photon wave function $  \psi_\lambda^{\gamma \to q \bar q}$ can be computed perturbatively~\cite{Kowalski:2006hc}, but the $\jpsi$ wave function $\psi_\lambda^{V\to q \bar q}$ is nonperturbative and we model it using the Boosted Gaussian ansatz~\cite{Kowalski:2006hc} with parameters from Ref.~\cite{Mantysaari:2018nng}.
We note that with the IP-averaged dipole amplitude we can only compute the  $t=0$ case
because the term $e^{-i\Deltat \vdot \bt}$ in Eq.~\eqref{eq:amplitude} makes the scattering amplitude sensitive to the shape of the target at finite $t$.
To compute the $t$-integrated cross section in this case, we take the shape of the $t$-distribution from the full IP-dependent case and write
\begin{equation}
    \int_0^\infty \dd{\abs{t}}  \frac{\dd{\sigma}}{\dd{t}} 
    =
     \frac{\dd{\sigma}}{\dd{t}} \Bigg|_{t=0}
     \times
    \qty[\sigma_\text{IP} \Bigg/ \qty(\frac{\dd{\sigma}_{\text{IP}}}{\dd{t}}) \Bigg|_{t=0}] \,.
\end{equation}
where $\sigma_\text{IP}$ is the result obtained by using the IP-dependent dipole amplitude.
Consequently, we are then effectively comparing the results for the differential cross sections at $t=0$.

The results with and without the IP-dependence in the initial condition for the BK evolution of the dipole amplitude are shown in Fig.~\ref{fig:exclusive_jpsi_p_ip_vs_noip}.
\update{
With IP dependence the evolution is almost linear in this logarithmic plot within this energy range and consistent with the data.
This is similar to previous studies using the JIMWLK evolution~\cite{Mantysaari:2022sux}
and also very close to what is obtained using the linear BFKL evolution~\cite{Penttala:2024hvp}.
This demonstrates that the nonlinear effects are weak for exclusive $\jpsi$ production with proton targets in the currently accessible energy range.}

\update{
We also see that
neglecting the IP dependence in the evolution slows down the evolution quite drastically for proton targets.}
This can be understood from the different behavior of the nonlinear term:
with the IP dependence, the dipole amplitude approaches zero as the dipole size grows whereas without the IP dependence it approaches unity.
This means that the nonlinear term starts to suppress the contribution from large dipoles in the IP-independent case even when one would not expect saturation effects to be important. 
This might lead to an overestimation of the saturation effects when the IP-independent dipole amplitude is used in the evolution.
For processes where the relevant momentum scale in the process is larger, such as $\Upsilon$ production, the difference between the two approaches is expected to be smaller.
We note, however, that the parameters in our initial condition and in the evolution (running-coupling scale) were chosen such that the IP-dependent case agrees with the $\jpsi$ production data. Hence one should not interpret this figure as evidence that the energy dependence of the data cannot be described by the IP-independent case (see e.g. Ref.~\cite{Mantysaari:2022bsp} for a description of the HERA data with an IP-independent evolution).

For comparison, we also show the case with $m' = \SI{0}{GeV}$ in the evolution to demonstrate that the correct energy dependence can also be captured without IP dependence, at least for the lower energies (see also Appendix~\ref{app:DIS} for a comparison to the total DIS cross section data). 
We note that the value $m' = \SI{0}{GeV}$ was chosen as it happens to coincide with the energy dependence of the data for lower energies.
Further tuning the saturation scale in the initial condition is expected to bring the IP-independent case even closer to the data.
The main takeaway of this figure is that ignoring the IP dependence in the evolution can lead to vastly different results, questioning the validity of this approximation for protons.
For Pb targets, the situation is quite different: now the results using both dipole amplitudes agree across the whole energy range accessible in current experiments. 
This is due to the infinite-target approximation being more accurate for heavy nuclear targets.

\subsection{Effect of the running-coupling prescription}
\label{sec:results_coupling}

We can also quantify the effect of the different running-coupling prescriptions in the evolution.
In Fig.~\ref{fig:exclusive_jpsi_alphas}, we show predictions for
exclusive $\jpsi$ photoproduction off proton and Pb targets,
using the different choices for the running of the coupling from Sec.~\ref{sec:running_coupling}.
We note that the daughter, parent, and smallest dipole prescriptions lead to similar results for the evolution whereas the Balitsky prescription has a much slower evolution speed. This is consistent with the original numerical analysis of different running-coupling prescriptions presented in Ref.~\cite{Albacete:2007yr} without including the IP dependence.
While the Balitsky prescription agrees with the smallest dipole when one of the dipoles is much smaller than the others, the differences between the two prescriptions are numerically important.

We can also study the effects of the running-coupling prescription on exclusive vector meson production off heavy nuclei once the energy dependence has been fixed to the proton data.
As the Balitsky prescription leads to a much slower energy dependence, we choose $\lambdaQCD = \SI{0.1}{GeV}$ in Eq.~\eqref{eq:strong_coupling} to match the proton data when this prescription is used.
For the other running-coupling prescriptions we keep the original $\lambdaQCD = \SI{0.025}{GeV}$ as all of them approximately yield the correct energy dependence.
With the energy dependence fixed to the proton data, the results for Pb targets shown in Fig.~\ref{fig:exclusive_jpsi_alphas_Pb} with different running-coupling prescriptions are very close to each other.
This is also true for the nuclear suppression factor shown in Fig.~\ref{fig:exclusive_jpsi_alphas_R}, defined as
\begin{equation}
\label{eq:R_A}
     R_A = \sqrt{\frac{\sigma^A}{\sigma^A_\text{IA}}} \,,
\end{equation}
where
\begin{equation}
\label{eq:sigma_IA}
    \sigma^A_\text{IA} 
    = \frac{\dd{\sigma^{\gamma + p \to V + p}}}{\dd{t}} \Bigg|_{t=0} \times \int_0^\infty \dd{\abs{t}}\abs{F(t)}^2 
\end{equation}
refers to the nuclear cross section using the impulse approximation.
The integrated form factor $ \int \dd{\abs{t}} \abs{F(t)}^2$
is computed following the CMS collaboration~\cite{CMS:2023snh}.
We see that the nuclear suppression factor is not sensitive to the running-coupling prescription used once the energy dependence of the dipole amplitude is fixed to the proton data.

In particular, we find that all running-coupling prescriptions result in too weak nuclear suppression. This manifests itself  as the $\gamma+\mathrm{Pb}\to\jpsi+\mathrm{Pb}$ cross section (Fig.~\ref{fig:exclusive_jpsi_alphas_Pb}) being overestimated at $W\gtrsim \SI{100}{GeV}$, or equivalently, the nuclear modification factor $R_A(\jpsi)$ (Fig.~\ref{fig:exclusive_jpsi_alphas_R}) 
having a too weak energy dependence. This is consistent with previous saturation model calculations generically overestimating the cross sections measured in ultra-peripheral collisions at the LHC~\cite{Lappi:2013am,Mantysaari:2017dwh,Luszczak:2019vdc,Sambasivam:2019gdd,Bendova:2020hbb,Mantysaari:2022sux,Mantysaari:2023xcu}, and confirms that this is a generic feature of the applied model and not due to a specific choice for the running-coupling prescription.

\subsection{Effect of the resummation of higher-order corrections}
\label{sec:results_constraint}

\begin{figure}[t!]
    \centering
    \begin{subfigure}[T]{0.9\columnwidth}
        \centering
        \includegraphics[width=\textwidth]{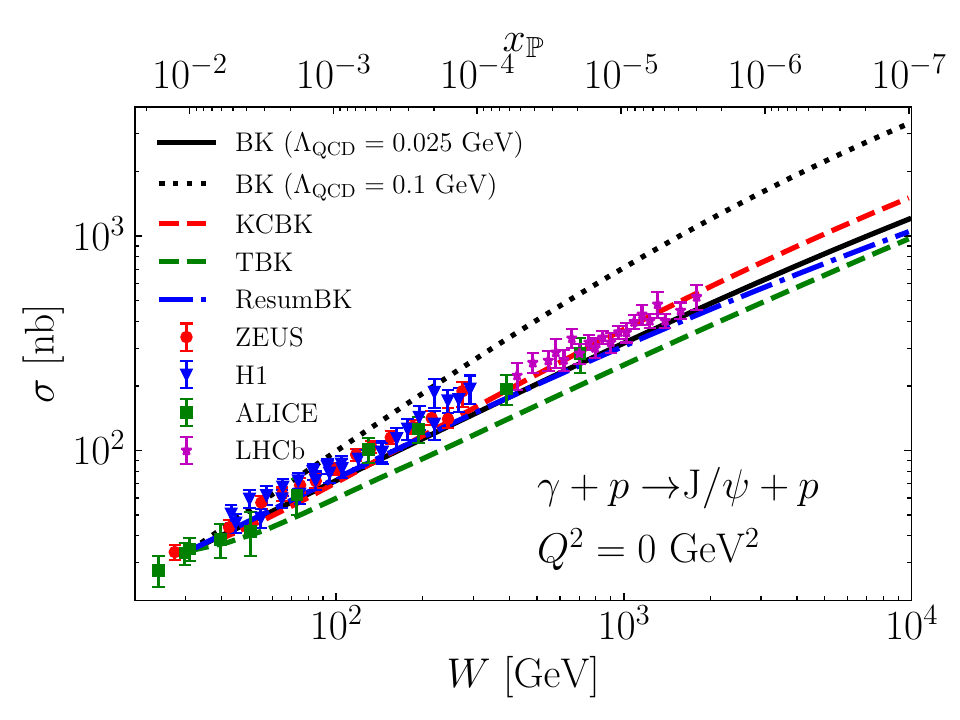}
        \caption{ Proton. }
        \label{fig:NLL_comparison_p}
    \end{subfigure}
    \begin{subfigure}[T]{0.9\columnwidth}
        \centering
        \includegraphics[width=\textwidth]{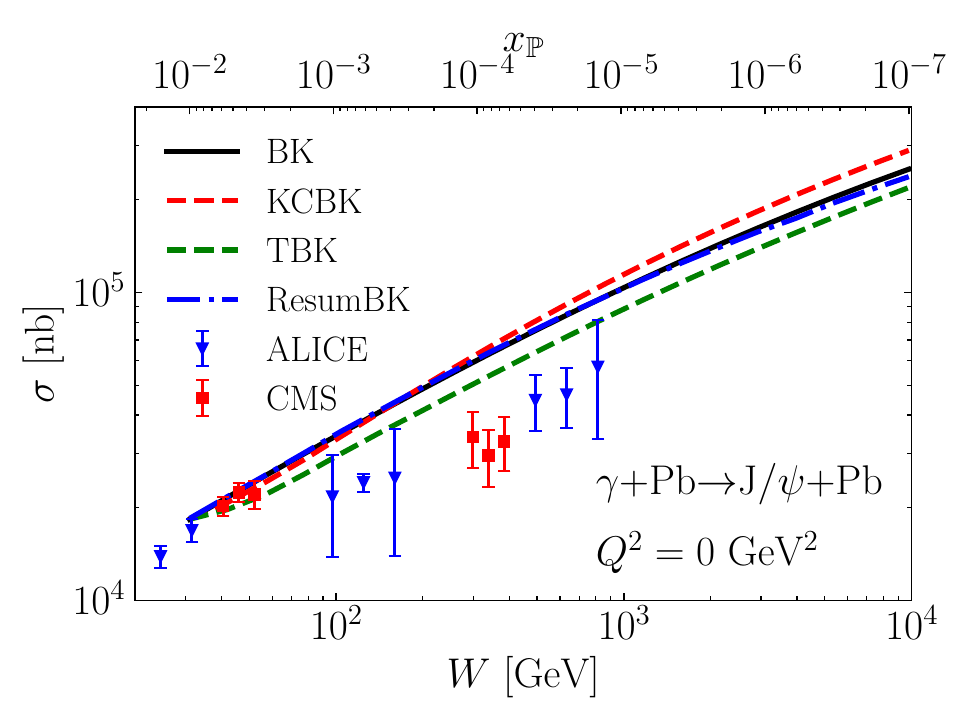}
        \caption{ Lead. }
        \label{fig:NLL_comparison_Pb}
    \end{subfigure}
    \begin{subfigure}[T]{0.9\columnwidth}
        \centering
        \includegraphics[width=\textwidth]{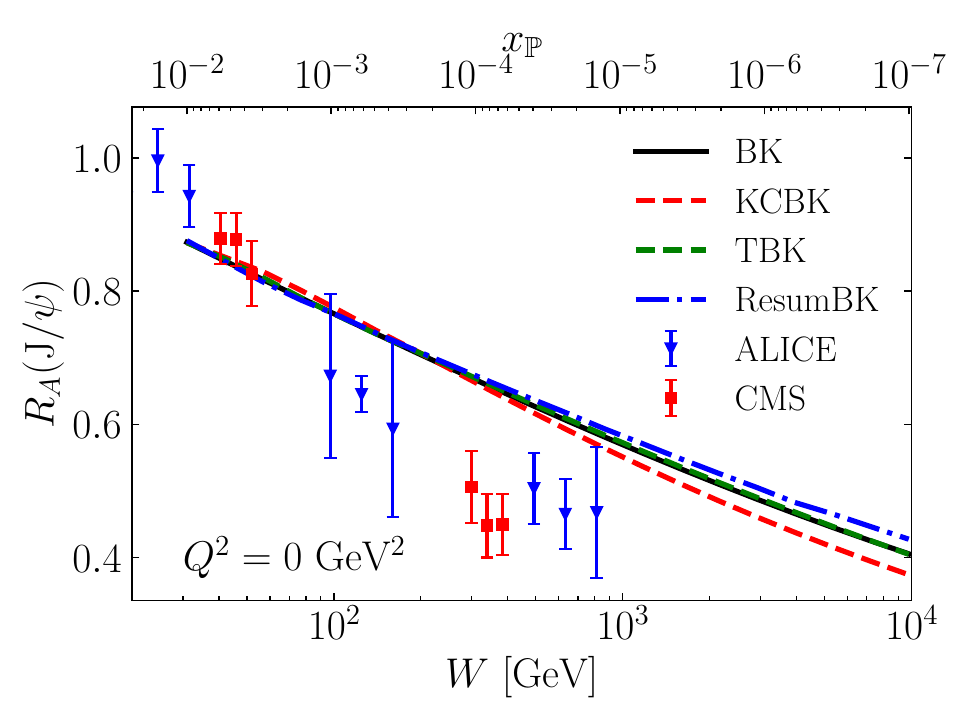}
        \caption{ Nuclear suppression factor.}
        \label{fig:NLL_comparison_R}
    \end{subfigure}
    \caption{
    Exclusive $\jpsi$ production data with
    different prescriptions of the resummation of higher-order corrections in the BK equation. The running-coupling scale parametrized by $\lambdaQCD$
     has been modified to match the energy dependence of the proton data.
    }
    \label{fig:NLL_comparison}
\end{figure}
To determine the effect of resumming large transverse logarithms that appear at higher orders in perturbation theory, we calculate exclusive $\jpsi$ photoproduction using the  three different resummation schemes discussed in Sec.~\ref{sec:kinematic_constraint}.
The results are shown in Fig.\,\ref{fig:NLL_comparison} where we have also chosen to use
$\lambdaQCD = \SI{0.1}{GeV}$ with all of the resummation schemes, to agree with the energy dependence of the proton data.
The resummation of higher-order corrections generally slows down the evolution, which explains the need for a larger value of $\lambdaQCD$ (see also Refs.~\cite{Kovchegov:2006vj,Lappi:2012vw} for a discussion of the preferred choice of the coordinate-space running-coupling scale).
We also note that while the TBK results agree with the energy dependence of the data, the overall normalization is off. 
This results from the shift going from the projectile to the target rapidity, which
means that the probed values of $\xpom$ are generally larger for TBK.
The rapidity shift mostly affects the overall normalization but not the energy dependence, and thus the effects of the shift are canceled in the nuclear suppression factor (Fig.~\ref{fig:NLL_comparison_R}).
Again, we see that once the energy dependence is fixed to the proton data, the predictions for the nuclear cross section (Fig.~\ref{fig:NLL_comparison_Pb}) and for the nuclear suppression factor (Fig.~\ref{fig:NLL_comparison_R}) are mostly independent of the resummation scheme. Similarly as in the case of different running-coupling prescriptions, this suggests that the challenges in simultaneously describing the $\gamma+p$ and $\gamma+\mathrm{Pb}$ data in previous CGC calculations are not due to the lack of the resummation of the large transverse logarithms.

\begin{figure}[t]
    \centering
    \includegraphics[width=\columnwidth]{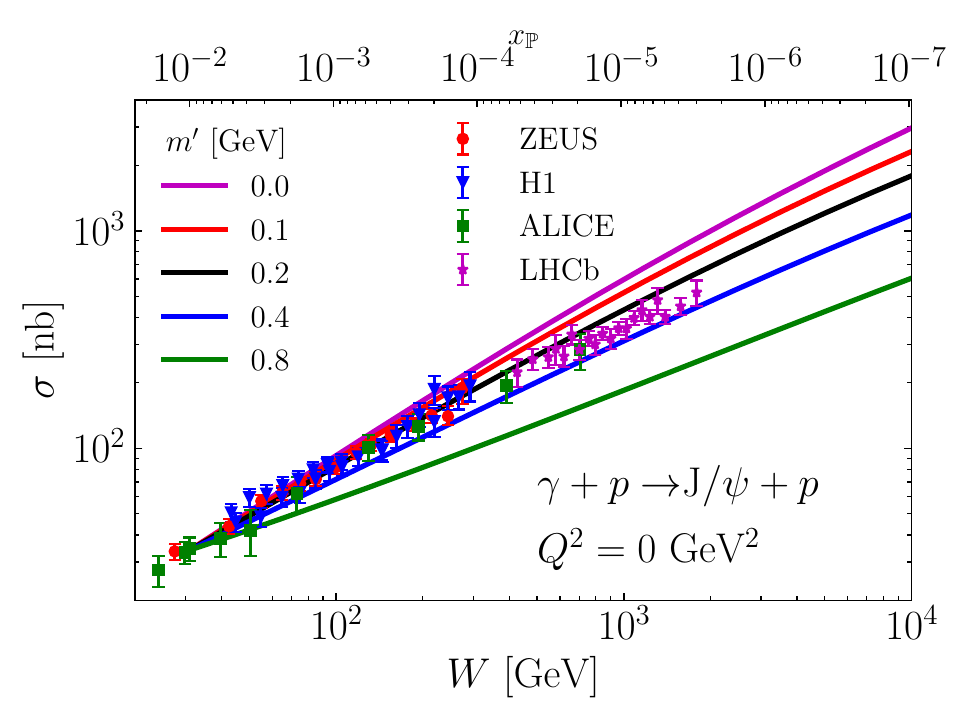}
    \caption{Exclusive $\jpsi$ production as a function of the center-of-mass energy $W$ for proton targets, shown for different values of the IR regulator $m'$ in the BK evolution.
    }
    \label{fig:exclusive_jpsi_m_dependence}
\end{figure}

\begin{figure}[t]
    \centering
    \includegraphics[width=\columnwidth]{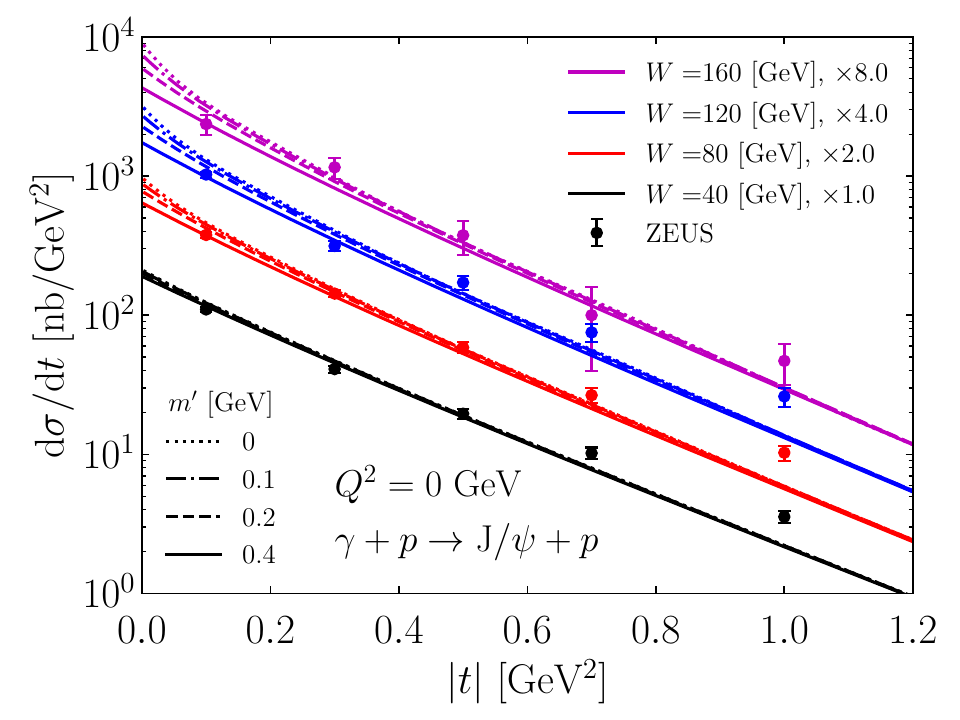}
    \caption{Differential cross section for exclusive $\jpsi$ production as a function of the momentum transfer $t$, compared to the ZEUS data~\cite{ZEUS:2002wfj}.
    Different linestyles correspond to different values of $m'$ in the evolution.
    To improve readability, results for different center-of-mass energies $W$ have been scaled by the factor shown in the figure.
    }
    \label{fig:exclusive_jpsi_dsigmadt}
\end{figure}

\subsection{Effect of the infrared regulator in the evolution}

\label{sec:regulator}

The BK evolution of an IP-dependent initial condition is generally sensitive to the value of the infrared regulator $m'$ in the evolution.
This is demonstrated
in Fig.~\ref{fig:exclusive_jpsi_m_dependence} where we show the results for exclusive $\jpsi$ production with proton targets, varying the infrared regulator $m'$ from our default setup $m'= \SI{0.4}{GeV}$.
The source of the sensitivity to the infrared regulator can be seen
in Fig.~\ref{fig:exclusive_jpsi_dsigmadt} where we plot the differential cross section for exclusive $\jpsi$ production as a function of the squared momentum transfer $t$ for various energies $W$.
Changing the value of $m'$ mostly affects the differential cross section for small values $t$ whereas the effect at larger $|t|$ is small.
This is because the infrared regulator $m'$ mostly controls the growth of the target that is sensitive to the evolution at long distance scales, and for large $|t|$ one is mostly sensitive to the target structure at short length scales $\sim 1/\sqrt{|t|}$.

\begin{figure}[t!]
    \centering
     \begin{subfigure}[T]{0.9\columnwidth}
        \centering
        \includegraphics[width=\textwidth]{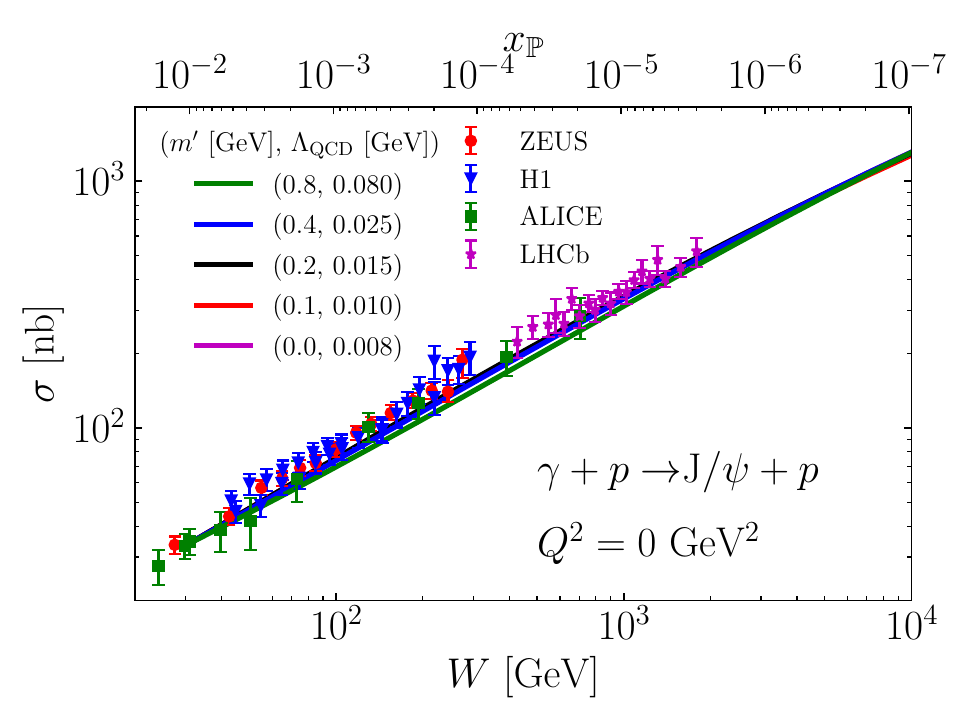}
        \caption{ Proton. }        \label{fig:exclusive_jpsi_m_dependence_adjusted_p}
    \end{subfigure}
    \begin{subfigure}[T]{0.9\columnwidth}
        \centering
        \includegraphics[width=\textwidth]{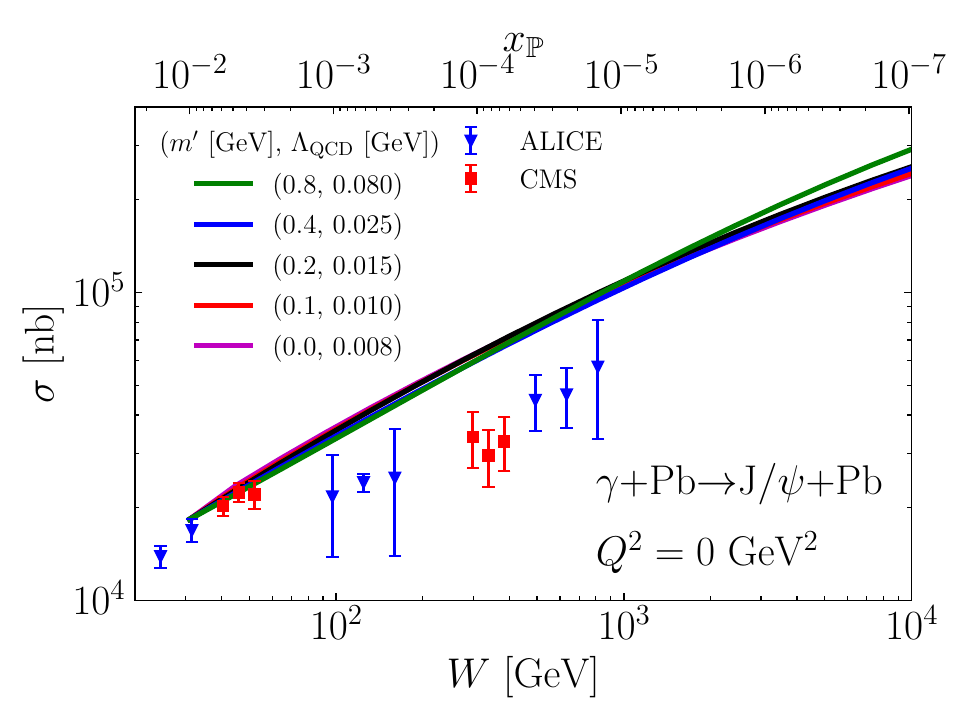}
        \caption{ Lead. } \label{fig:exclusive_jpsi_m_dependence_adjusted_Pb}
    \end{subfigure}
    \begin{subfigure}[T]{0.9\columnwidth}
        \centering
        \includegraphics[width=\textwidth]{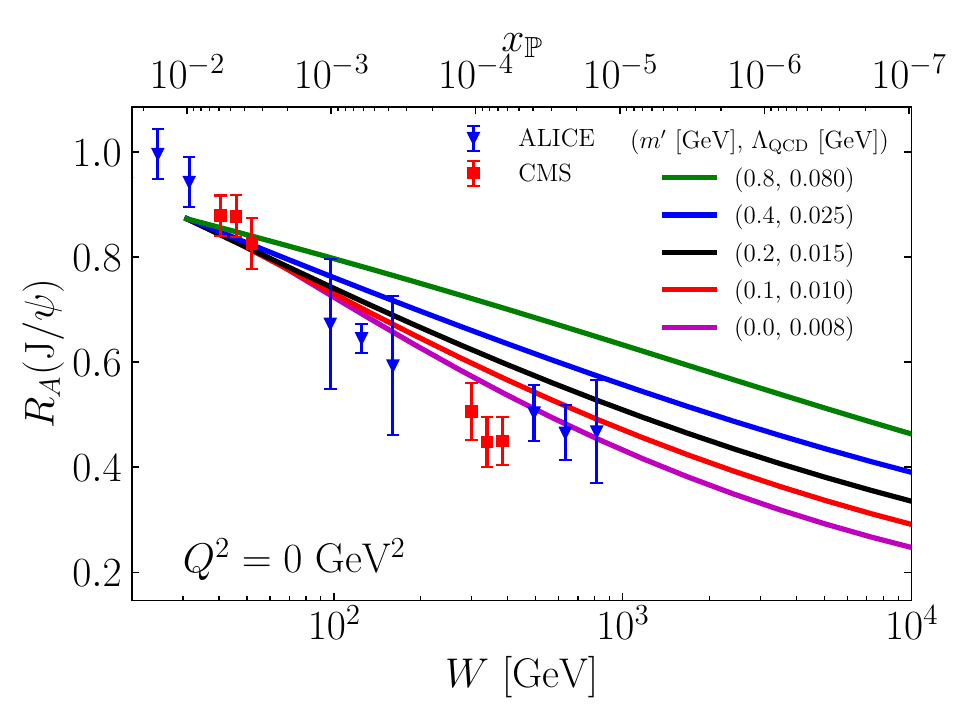}
        \caption{ Nuclear suppression factor.}
        \label{fig:exclusive_jpsi_m_dependence_adjusted_R}
    \end{subfigure}
    \caption{Exclusive $\jpsi$ production as a function of the center-of-mass energy $W$, shown for different values of the IR regulator $m'$ in the BK evolution.
    The constant $\lambdaQCD$ in $\as$ has been adjusted to match the data.
    }
    \label{fig:exclusive_jpsi_m_dependence_adjusted}
\end{figure}

However, we note that we can compensate for the sensitivity of the integrated cross section and its $W$ dependence to $m'$ by also changing the value of $\lambdaQCD$ in the coupling constant (Eq.~\eqref{eq:strong_coupling}).
This is shown in Fig.~\ref{fig:exclusive_jpsi_m_dependence_adjusted_p} where we consider different values of $m'$ but also fix the parameter $\lambdaQCD$  to the energy dependence of the proton data.
Using these values of $m'$ and $\lambdaQCD$ that have been fixed to the proton data, we can then compute results for nuclear targets.
This is shown in Fig.~\ref{fig:exclusive_jpsi_m_dependence_adjusted_Pb} where we see that the nuclear results do not change much as long as the evolution agrees with the proton data.
In contrast, the value  of $m'$ has a huge impact on the nuclear suppression factor as demonstrated in Fig.~\ref{fig:exclusive_jpsi_m_dependence_adjusted_R}
where we show the nuclear suppression factor with the same values of $m'$ and $\lambdaQCD$.
This is because the nuclear suppression factor is defined using the proton differential cross section at $t=0$ which is very sensitive to the value of $m'$ as we have already noted.

\update{
While the nuclear suppression factor data seemingly prefers $m'=0$, we note that the experimental data
requires the value of $\dd\sigma/\dd t$  in $\gamma+p$ at $t=0$ which cannot be measured. 
Instead, it has been computed by extrapolating the measured $t$ spectrum with an exponential form $\exp(-B\abs{t})$, where $B$ is a (energy-dependent) constant fitted to the HERA data.
However, 
this functional form is not compatible with our theoretical predictions
for small values of $m'$ as can be seen
from Fig.~\ref{fig:exclusive_jpsi_dsigmadt},
where this exponential form would correspond to a straight line.
As such, the procedure to obtain the nuclear modification factors with small $m'$ is not consistent with the experimental method.}
This is also reflected in the fact that all of these setups give essentially identical results for the integrated cross section and only the nuclear suppression results differ.
As such, we advocate that the nuclear modification factor measurements should be made less model-dependent by directly reporting the $t$-integrated cross section ratio with a proper scaling as previously suggested e.g. in Ref.~\cite{Mantysaari:2018nng}.

{We also note that the applied setup cannot describe the ZEUS data at high $|t|\sim \SI{1}{GeV^2}$. In principle this could be seen to suggest that a different model for the proton geometry should be used~\cite{Goncalves:2024jlx}. However, we also note that when a similar setup is compared to H1 data~\cite{H1:2013okq} in Ref.~\cite{Mantysaari:2022sux}, a good description of the H1 measurement over a comparable $t$ range is obtained. As such we consider the description of the proton geometry to be realistic, and emphasize that the conclusions of this work are not expected to be sensitive to the detailed shape assumed for the proton or the Pb nucleus. }

\section{Conclusions}
\label{sec:conclusions}

In this work, we have studied the importance of including the full impact-parameter dependence in the small-$x$ evolution of the dipole amplitude.
This has been done by considering a generalization of the McLerran--Venugopalan model that includes the dependence on the transverse coordinate in the color correlator, and then evolving this dipole amplitude to higher energies with the BK equation.
This IP-dependent setup is compared to the case where 
we first integrate over the impact parameter before the evolution, effectively employing an IP-averaged dipole amplitude.
It can be argued that this corresponds to neglecting the dependence on the impact parameter, as is usually done in phenomenological applications.
We have demonstrated that this approximation is generally not applicable for
protons, leading to an overestimation of saturation effects.
For lead targets, on the other hand, neglecting the IP dependence has a much smaller effect, which follows from the larger size of the lead nucleus.

The difference between evolution for initial conditions with and without the IP dependence comes from the nonlinear term in the BK equation responsible for saturation effects.
When we ignore the IP dependence, the dipole amplitude approaches unity for large dipoles and thus the nonlinear term becomes increasingly important.
When the IP dependence is taken into account, the effect is the opposite:
for large dipoles the dipole amplitude becomes negligible, making the nonlinear term relatively more suppressed compared  to the linear terms.
This also means that with the full IP dependence the nonlinear term does not suppress the contribution from 
large dipoles, 
making it more sensitive to gluon splitting at nonperturbative scales.
Because of this,
we need to introduce an additional description of the nonperturbative physics { in the form of an infrared regulator} to get the correct energy dependence of the cross sections.

{
The energy dependence of the small-$x$ evolution can also be affected by higher-order corrections and the running-coupling prescription.
These modifications are difficult to implement in the Langevin formulation of the JIMWLK evolution (with which we find good agreement, see Appendix~\ref{app:JIMWLK}), but can be studied in the BK evolution, and we do so here with full IP dependence.}
One motivation for looking into the effect of the modified BK evolution is the observation that the CGC approach typically leads to an overestimation of the exclusive $\jpsi$ production cross section for nuclear targets.
We have demonstrated that neither the resummation of higher-order corrections nor a choice of a different running-coupling prescription allow for a simultaneous description of exclusive $\jpsi$ production in $\gamma+p$ and $\gamma+$Pb collisions. 
{
Similarly, once the energy dependence of the cross section is fixed to the data for proton targets, the effect of varying the infrared regulator is negligible for Pb targets.
However, we note that even then the $t$-differential cross section can be sensitive to the value of the IR regulator at small $t$.
This makes the nuclear suppression factor in its current definition (Eq.~\eqref{eq:R_A}) very sensitive to the extrapolation of the differential cross section down to $t=0$, and we encourage using the directly measurable $t$-integrated cross section instead in defining this ratio.
}

In the future, it will be desirable to constrain the IP-dependent initial state by considering both inclusive and diffractive processes.
Our current setup, constrained solely by exclusive $\jpsi$ production data, fails to provide the correct normalization for inclusive DIS as shown in Appendix~\ref{app:DIS}.
Of course, diffractive processes are essential for constraining IP dependence, as the $t$-dependent cross section carries information on the target shape, which cannot be probed in inclusive processes.
On the other hand, inclusive DIS can be measured very accurately and is not sensitive to the geometry that depends on non-perturbative physics. {Consequently, it can provide reliable constraints on the dipole amplitude. A global fit to all available data, including the full IP dependence, is expected to provide better constraints on the model parameters. }

\begin{acknowledgments}
    We thank Paul Caucal, Edmond Iancu, Alexander Kovner, Tuomas Lappi, Yacine Mehtar-Tani, Christophe Royon, and Raju Venugopalan for discussions.
    J.P is supported by the National Science Foundation under grant No. PHY-1945471, and by the U.S. Department of Energy, Office of Science, Office of Nuclear Physics, within the framework of the Saturated Glue (SURGE) Topical Theory Collaboration.
    H.M is supported by the Research Council of Finland, the Centre of Excellence in Quark Matter and projects 338263, 346567 and 359902, and by the European Research Council (ERC, grant agreements ERC-2023-101123801 GlueSatLight and ERC-2018-ADG-835105 YoctoLHC). F.S. is supported by the Institute for Nuclear Theory’s U.S. DOE under Grant No. DE-FG02-00ER41132. The content of this article does not reflect the official opinion of the European Union and responsibility for the information and views expressed therein lies entirely with the authors. Computing resources from CSC – IT Center for Science in Espoo, Finland and from the Finnish Computing Competence Infrastructure (persistent identifier \texttt{urn:nbn:fi:research-infras-2016072533}) were used in this work. This material is based upon work supported by the U.S. Department of Energy, Office of Science, Office of Nuclear Physics, under DOE Contract No.~DE-SC0012704. 
\end{acknowledgments}

\appendix

\section{Comparions to JIMWLK}
\label{app:JIMWLK}

In the Langevin form, which is suitable for numerical implementation~\cite{Weigert:2000gi,Blaizot:2002np,Dumitru:2011vk,Lappi:2012vw,Cali:2021tsh}, the JIMWLK equation can be written as a stochastic evolution equation for the Wilson lines $V(\xt)$. In the form obtained in Ref.~\cite{Lappi:2012vw}, the evolution equation can be written as
\begin{multline}
\label{eq:jimwlk}
V_\xt(y+\der y) = \exp \left\{ -i \frac{\sqrt{\as \der y}}{\pi} \int \der^2 \zt\, \mathbf{K}_{\xt-\zt} \cdot (V_\zt \boldsymbol{\xi}_\zt V^\dagger_\zt) \right\}\\
\times V_\xt(y) \exp \left\{ i\frac{\sqrt{\as \der y}}{\pi} \int \der^2 \zt\, \mathbf{K}_{\xt-\zt} \cdot \boldsymbol{\xi}_\zt \right\},
\end{multline}
where $\boldsymbol{\xi}_\zt = (\boldsymbol{\xi}_{\zt,1}^a t^a, \boldsymbol{\xi}_{\zt,2}^a t^a)$, and the $\boldsymbol{\xi}_{\zt,i}^a$ describe local random Gaussian noise. The JIMWLK kernel $\mathbf{K}_{\xt}$, describing the probability density for the gluon emission, is given in Eq.~\eqref{eq:BK_kernel_regulator}.

We solve the JIMWLK evolution using the IP-MV model, specified in Sec.~\ref{sec:initial_condition}, as an initial condition. The numerical implementation of the MV model is based on the public IP-Glasma code~\cite{ipglasma_code}. The JIMWLK evolution is performed on a $\SI{5.12}{fm^2}$ lattice with $2048^2$ Wilson lines using the public code~\cite{jimwlk_code}. This setup is identical to the one used e.g.~in Refs.~\cite{Mantysaari:2022sux,Mantysaari:2023xcu}.

\begin{figure}[t]
    \centering
    \includegraphics[width=\columnwidth]{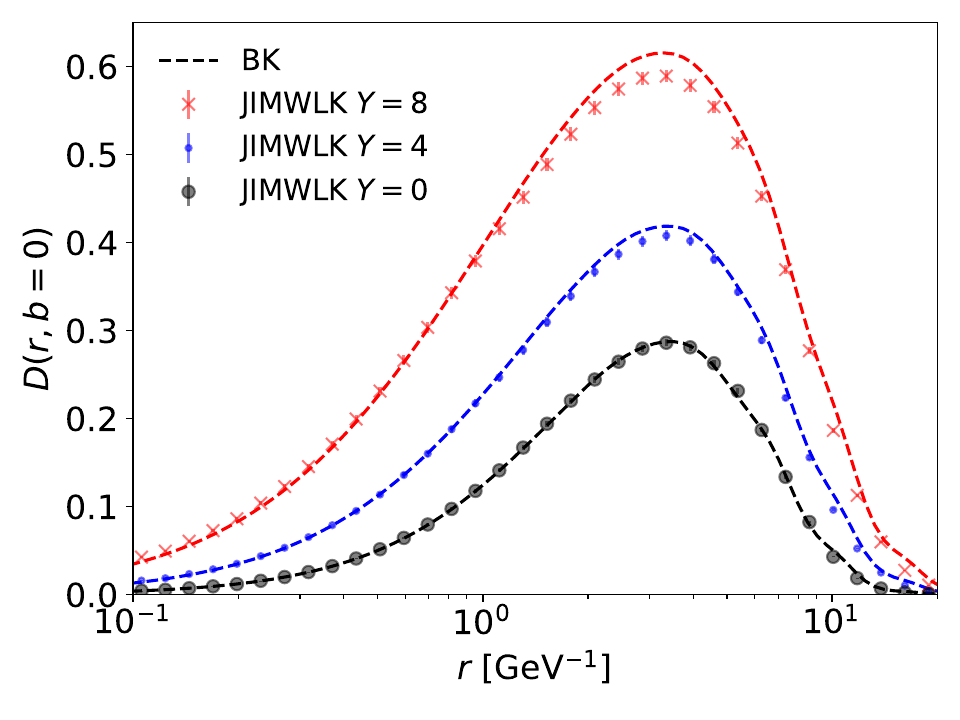}
    \caption{
    Comparison of the dipole amplitude evolved with the JIMWLK and BK equations.
    The dipole amplitude is shown as a function of the dipole size $r$ with the fixed impact parameter $b=0$ for different rapidities.
    }
    \label{fig:ipmv_lattice_jimwlk_r}
\end{figure}

\begin{figure}[t]
    \centering
    \includegraphics[width=\columnwidth]{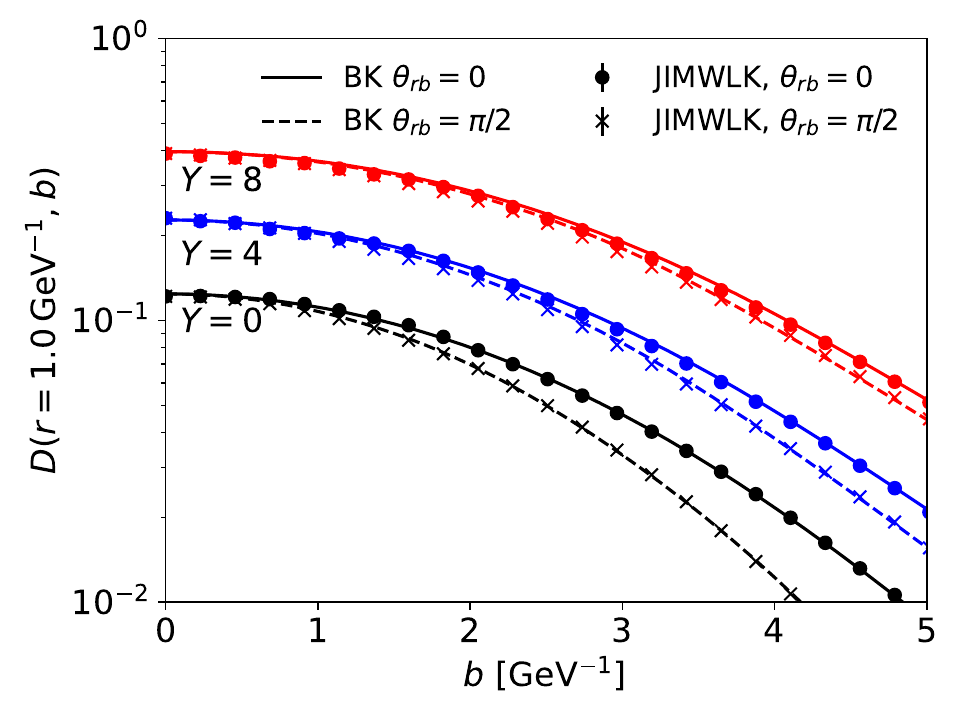}
    \caption{
    IP dependence of the dipole amplitude, from the BK evolution of an  IP-dependent  initial condition compared to the lattice MV model + JIMWLK. The results are shown separately for the case where the dipole is aligned parallel ($\theta_{rb}=0$) and perpendicular ($\theta_{rb}=\pi/2$) to the impact parameter.
    }
    \label{fig:ipmv_lattice_jimwlk_b}
\end{figure}

The dipole--proton scattering amplitudes as a function of dipole size $r$ and impact parameter $b$ are shown in Figs.~\ref{fig:ipmv_lattice_jimwlk_r} and~\ref{fig:ipmv_lattice_jimwlk_b}, respectively. The daughter-dipole running-coupling prescription is used in both setups.
Fig.~\ref{fig:ipmv_lattice_jimwlk_r} shows the dipole--proton scattering amplitude in the case where the center of the dipole $\bt$ is at the center of the proton (origin). 
The BK and JIMWLK evolution result in very similar dipole amplitudes except at very large or small $r$ (compared to the lattice size) where the lattice artifacts dominate.  
The good agreement is expected, as although the BK equation is obtained only in the mean-field limit $\langle \hat S \hat S\rangle \approx \langle \hat S \rangle \langle \hat S \rangle$ from the JIMWLK evolution, which can be seen to follow from the large-$\Nc$ limit, the difference between the JIMWLK and BK equations is known to be much smaller than the naive expectation $\sim 1/\Nc^2\sim 10\%$~\cite{Kovchegov:2008mk}. However, we note that a slightly smaller evolution speed is obtained with the JIMWLK evolution, with a few percent differences seen in the dipole amplitudes at $Y=8$. This observation is consistent with Refs.~\cite{Lappi:2012vw,Kovchegov:2008mk}. A more quantitative comparison of these minor differences would require a more detailed continuum extrapolation that goes beyond the scope of this work.  

Similarly, a good agreement between the dipole amplitudes as a function of impact parameter $\bt$ can be seen in Fig.~\ref{fig:ipmv_lattice_jimwlk_b}. Here, two different orientations of the dipole relative to the impact parameter direction are shown: either the dipole is aligned parallel ($\theta_{rb}=0$) or perpendicular ($\theta_{rb}=\pi/2$) to the impact parameter. The scattering amplitude is larger for a dipole that is parallel to the impact parameter in the region where there is a large density gradient (see also discussion in Ref.~\cite{Mantysaari:2019csc}). The JIMWLK and BK evolutions are found to result in identical impact-parameter dependencies for both orientations.

\begin{figure}[t]
    \centering
    \includegraphics[width=\columnwidth]{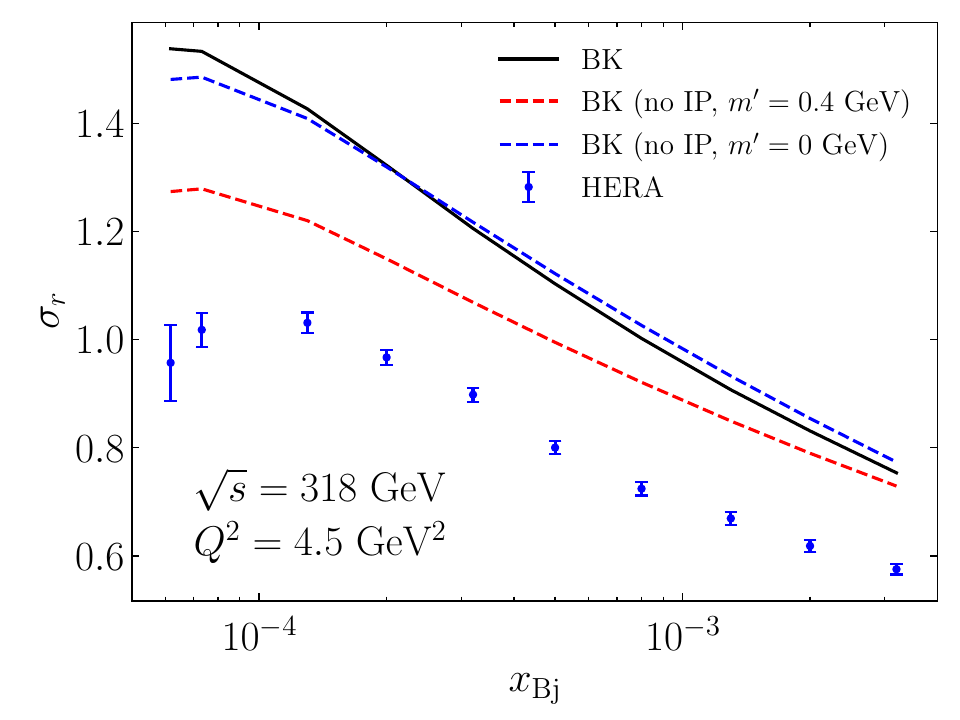}
    \caption{The reduced cross section $\sigma_r$ as a function of $\xbj$, with and without IP dependence in the initial condition for the evolution, 
    compared to the HERA data from Ref.~\cite{H1:2009pze}. 
    }
    \label{fig:F2_ip_vs_noip}
\end{figure}

\section{Reduced cross section in DIS}
\label{app:DIS}

In this Appendix, we consider the reduced cross section in DIS in the kinematics of the HERA collider
with the same setup for the dipole amplitudes as in Sec.~\ref{sec:IPvsNOIP}.
The reduced cross section is computed at leading order~\cite{Kowalski:2006hc,Albacete:2010sy,Lappi:2013zma}, and
we
take into account both light quarks ($u,d,s$) with $m_q=0$ and the charm quark with $m_q=\SI{1.3528}{GeV}$, consistent with our setup for $\jpsi$ production. 
The results are shown in Fig.~\ref{fig:F2_ip_vs_noip} for the photon virtuality $Q^2=\SI{4.5}{GeV^2}$.
The overall normalization of the results does not agree with the experimental data for inclusive DIS, which is due to the fact that the initial condition was constrained by exclusive $\jpsi$ production data but not by inclusive DIS.
This tension is in line with the previous observations of Ref.~\cite{Mantysaari:2018zdd}.
The overall normalization of exclusive $\jpsi$ production is sensitive to many factors such as the exact value of the charm quark mass and the chosen model for the vector meson wave function~\cite{Kowalski:2006hc,Lappi:2020ufv} which is one possible explanation for the discrepancy.
To make more consistent comparisons to the inclusive DIS data, one should include it in constraining the initial condition.

In this work, we are more interested in seeing the effects of including the IP dependence in the evolution.
While this is somewhat sensitive to the overall normalization, in the sense that the normalization is affected by the saturation scale, the qualitative features revealed by comparing different setups should not depend on it strongly.
Comparing the two setups with and without the IP dependence in the initial condition for the BK evolution, we see a large difference between the results after evolution at the total cross section level. In particular, the  evolution of the IP-independent initial condition is found to be slower.
This is consistent with the results for exclusive $\jpsi$ production shown in Sec.~\ref{sec:IPvsNOIP}.

Again, we also show the results for the evolution without the IP dependence with $m'= 0$ which happens to coincide with our results for the evolution of the IP-dependent initial condition.
For the values of $\xbj$ probed at HERA, the differences between the evolution of the IP-dependent and IP-independent initial condition with $m'=0$ are too small to distinguish the two models based on the experimental data.
Thus, one would not expect to see problems of ignoring the IP dependence in fits of the dipole amplitude to the structure function data only.
This is in contrast to exclusive $\jpsi$ production where the deviations can already be seen in the kinematical region covered by the currently available data.

\bibliographystyle{JHEP-2modlong.bst}
\bibliography{ref}

\end{document}

%% file: packages.tex
\usepackage{graphicx}
\usepackage{bm,amsmath,amssymb}
\usepackage{xcolor}
\usepackage{mathrsfs}
\usepackage{mathtools}
\usepackage{slashed}
\usepackage{physics}
\usepackage{siunitx}

\usepackage{float}

\definecolor{lcolor}{rgb}{0.5,0,0}
\definecolor{citcolor}{rgb}{0,0.3,0.0}

\usepackage[breaklinks,colorlinks,urlcolor=blue,citecolor=citcolor,linkcolor=lcolor]{hyperref}

\usepackage{subcaption}

%% file: definitions.tex
\newcommand{\as}{\alpha_{\mathrm{s}}}

\newcommand{\vect}[1]{\mathbf{#1}} 

\newcommand{\bt}{\vect{b}}

\newcommand{\Kt}{\vect{K}}

\newcommand{\xt}{\vect{x}}
\newcommand{\yt}{\vect{y}}
\newcommand{\zt}{\vect{z}}

\newcommand{\rt}{\vect{r}}

\newcommand{\jpsi}{\text{J}/\psi}

\newcommand{\Deltat}{\boldsymbol{\Delta}}
\newcommand{\xpom}{x_{\mathbb{P}}}
\newcommand{\Ypom}{Y_{\mathbb{P}}}
\newcommand{\xbj}{x_\text{Bj}}

\newcommand{\Yif}{Y_{0,\text{BK}}}
\newcommand{\etaif}{\eta_{0,\text{BK}}}
\newcommand{\lambdaQCD}{\Lambda_\text{QCD}}

\newcommand{\gs}{g_\mathrm{s}}
\newcommand{\Nc}{N_c}
\newcommand{\der}{\mathrm{d}}